\documentclass[review,onefignum,onetabnum]{siamonline190516}

\usepackage{lipsum}
\usepackage{amsfonts}
\usepackage{graphicx}
\usepackage{epstopdf}
\usepackage{algorithmic}
\usepackage{subfig}
\usepackage{rotating,booktabs,multirow}
\usepackage{amsmath,bm}

\ifpdf
  \DeclareGraphicsExtensions{.eps,.pdf,.png,.jpg}
\else
  \DeclareGraphicsExtensions{.eps}
\fi

\newcommand{\bx}{\mathbf{x}}
\newcommand{\bk}{\mathbf{k}}
\newcommand{\ba}{\mathbf{a}}
\newcommand{\dontshow}[1]{}

\def\kwave~{\textbf{k}-\texttt{Wave}}

\newcommand{\todo}[1]{\textcolor{red}{#1}}

\newcommand{\rv}[1]{\textcolor{blue}{#1}}
\newcommand{\bp}[1]{\textcolor{magenta}{#1}}

\DeclareFontEncoding{LS1}{}{}
\DeclareFontSubstitution{LS1}{stix}{m}{n}

\DeclareRobustCommand{\rightangle}{%
  \text{\usefont{LS1}{stix}{m}{n}\symbol{"D9}}%
}

\DeclareMathOperator*{\argminA}{arg\,min}

\makeatletter
\newenvironment{VR-FISTA}[1][htb]{%
    \renewcommand{\ALG@name}{VR-FISTA}
   \begin{algorithm}[#1]%
  }{\end{algorithm}}
\makeatother

\makeatletter
\newlength\tmp@\newlength\t@mp
\newcommand{\comp}[3]
  {\mathop{ \settowidth\tmp@{$\displaystyle\mathop{#1}^{#3}_{#2}$}
  \hbox to \tmp@{\hss \settowidth\t@mp{$\displaystyle #1$}\setlength\t@mp{.45\t@mp}
  $\displaystyle\mathop{#1}^{\hspace\t@mp #3}_{\hspace{-\t@mp}#2}$
  \hss} }}
\makeatother

\usepackage{enumitem}
\setlist[enumerate]{leftmargin=.5in}
\setlist[itemize]{leftmargin=.5in}

\newsiamremark{remark}{Remark}
\newsiamremark{hypothesis}{Hypothesis}
\crefname{hypothesis}{Hypothesis}{Hypotheses}
\newsiamthm{claim}{Claim}

\headers{On Learning the Invisible in limited angle PAT}{Bolin Pan and Marta M. Betcke}

\title{On Learning the Invisible in Photoacoustic Tomography with Flat Directionally Sensitive Detector\thanks{Submitted to the editors April 16, 2022.
\funding{M.~Betcke would like to acknowledge support from EPSRC EP/T014369/1, EP/W007673/1.}}}

\author{Bolin Pan\thanks{Department of Computer Science, University College London, UK
  (\email{bolin.pan.15@ucl.ac.uk}).}
  \and Marta M. Betcke\thanks{Department of Computer Science, University College London, UK
  (\email{m.betcke@ucl.ac.uk})}}

\usepackage{amsopn}
\DeclareMathOperator{\diag}{diag}

\DeclareMathOperator{\sign}{sign}

\ifpdf
\hypersetup{
  pdftitle={CurveNet},
  pdfauthor={Bolin Pan and Marta M.~Betcke}
}
\fi

\begin{document}
\nolinenumbers

\maketitle

\begin{abstract}
  In photoacoustic tomography (PAT) with flat sensor, we routinely encounter two types of limited data. The first is due to using a finite sensor and is especially perceptible if the region of interest is large relative to the sensor or located farther away from the sensor. In this paper, we focus on the second type caused by a varying sensitivity of the sensor to the incoming wavefront direction which can be modelled as binary i.e.~by a cone of sensitivity. Such visibility conditions result, in the Fourier domain, in a restriction of 
  both the image and the data to a bow-tie, akin to the one corresponding to the range of the forward operator. The visible wavefrontsets in image and data domains, are related by the wavefront direction mapping.
  We adapt the wedge restricted Curvelet decomposition, we previously proposed for the representation of the full PAT data, to separate the visible and invisible wavefronts 
  in the image.
  We optimally combine fast approximate operators with tailored deep neural network architectures into efficient learned reconstruction methods which perform reconstruction of the visible coefficients and the invisible coefficients are learned from a training set of similar data. 
\end{abstract}



\section{Introduction}
\label{sec:introduction}
Photoacoustic Tomography (PAT) is a hybrid imaging modality that can deliver high resolution \emph{in-vivo} images of optical energy absorbed upon an illumination with a laser-generated near infra-red light pulse \cite{wang2009multiscale, beard2011biomedical, nie2014structural, valluru2016photoacoustic, zhou2016tutorial, xia2013small}. PAT harnesses the photoacoustic effect to encode optical contrast onto broadband ultrasonic waves thereby circumventing the depth and spatial resolution limitations of purely optical imaging techniques.

If the data is measured over a surface surrounding the region of interest for long enough time and using omnidirectional sensors, the reconstruction can be obtained via \emph{time reversal} which amounts to a single solve of a wave equation; see \Cref{sec:PAT} for details. In practice, the limited-view problems are introduced due to spatially and/or directionally restricted sampling of the ultrasound (US) waves. An example of the former is a finite size linear/planar sensor placed on one side of the domain;  
the latter most US sensors' incapability of recording wavefronts impinging on the detector at angles beyond $\pm\theta_{\max}, \, \theta_{\max} < \pi/2$. Another type of limited-view problems result from restricting the length of the recorded time series, a necessity arising in a case of trapping sound speeds. As in this work we focus on constant speed of sound, we do not consider this last type of limited-view problems.

For limited-data problems due to sparse sampling rather than limited-view, compressed sensing (CS) approaches that iteratively minimise a penalty function combining an explicit model of US propagation 
with constraints on the image as regularisation \cite{huang2013full, arridge2016adjoint, arridge2016accelerated, haltmeier2016compressed, huynh2017sub, boink2017reconstruction, schwab2018galerkin, pan2021photoacoustic, frikel2018_VWPAT, frikelQuinto2015_limitedViewTomoPAT, frikel2010_limitedXray, frikel2013sparse, frikel2013_limitedXray, quinto1993singularities} have proven to provide significantly better reconstructions than \emph{time reversal} and more general Neumann series based iterative methods which lack inherent regularisation.
A shared drawback of all these methods is high computational complexity of iterating with the forward and adjoint/inverse operators. This shortcoming has been addressed by two step methods which reconstruct complete data first and then obtain the image through single time reversal \cite{haltmeier2016compressed, Acoustic, pan2021photoacoustic}, which however usually sacrifice some reconstruction quality for efficiency.

For the considered limited-view problems the situation is direr. Time reversal and Neumann series methods as expected only reconstruct the visible singularities producing images with characteristic limited-view artefacts and lack noise suppression properties of CS methods \cite{xu2004reconstructions}. Performance of compressed sensing depends on the choice of the prior, e.g.~while total variation (TV) yields overall better reconstructions than Curvelet sparsity (reasons for which are discussed in \Cref{sec:sec:ImperfectLearning}), even for TV the directions of singularities missing from the data result in inherent blur in the reconstructed images. 

A decade after the breakthrough in Deep Neural Networks, Deep Learning (DL) techniques are ubiquitous in tomographic imaging \cite{kang2017deep, jin2017deep, hammernik2018learning, adler2017solving, zhu2018image, adler2018learned, arridge2019solving}. In comparison to CS, the learned approaches i) need fewer applications of the forward/adjoint operator which leads to more efficient reconstruction algorithms, 
ii) have the ability to extract prior information about the image from a training set of similar images which makes them particularly suitable for e.g.~medical image reconstruction where patient studies fit such scenario. DL approaches to PAT image reconstruction mainly fall in one of the two categories: i) learned post-processing of a simple reconstruction such as e.g.~pseudo-inverse, ii) learned iterative reconstruction (also referred to as model based learning). 

The learned post-processing usually employs a network with many layers and learnable parameters capable of encoding complicated image priors. The learned post-processing of the universal back-projection \cite{xu2005universal} was studied in \cite{antholzer2019deep, antholzer2018deep, antholzer2018photoacoustic, schwab2018real, schwab2019learned, lan2019net}, of the first iteration of an averaged time reversal in \cite{shan2019accelerated} 
and of time reversal in \cite{guan2019fully}. All but the last work use standard U-Net architecture, while the last uses a dense U-Net (a U-Net with dense blocks and skip connections).
The main advantage of learned post-processing is that 
an initial simple reconstruction
can be decoupled from the iterative training process. This obviously limits the impact of the physical model on the reconstruction procedure and hence makes the process less robust and more dependent on the training set used.

Learned iterative reconstruction approaches are based on unrolling of iterative solvers and parametrising the proximal map with neural networks, usually much smaller than those used in learned post-processing, e.g.~\cite{shan2019simultaneous,boink2018sensitivity, boink2019partially, boink2019robustness,yang2019accelerated}. 
Due to repeated application of the forward and adjoint/approximate inverse operators on the training set, learned iterative reconstruction methods for 3D PAT suffer from unreasonable training times. To tackle this, a greedy training for learned gradient schemes for 3D PAT was suggested in \cite{hauptmann2018model}. The same authors later considered use of efficient Fourier domain approximate forward model to accelerate the training process \cite{hauptmann2018approximate}, which is also the direction we take in this work.
\dontshow{
In 2D, simultaneous initial pressure and sound speed reconstruction was presented in \cite{shan2019simultaneous} and learned primal dual framework for simultaneous initial pressure reconstruction and segmentation was investigated in a series of papers \cite{boink2018sensitivity, boink2019partially, boink2019robustness}. Also in 2D, a network architecture based on recurrent inference machines for PAT reconstruction with approximate forward operator was proposed in \cite{yang2019accelerated}. 
}

Other proposed approaches include learned pre-processing of the data prior to reconstruction, fully learned methods completely bypassing the physics of the model, and methods that aim to learn some representation of the regulariser for use in variational framework, we refer to \cite{hauptmann2020deep} for a review of such approaches specific to PAT.

\subsection{Related Work}
\label{sec:sec:relatedwork}
In a series of papers Frikel and Quinto \cite{quinto1993singularities, frikel2010_limitedXray, frikel2013_limitedXray, frikel2013sparse, quinto2017artifacts} characterised the visibility of singularities in X-ray tomography and in \cite{xu2004reconstructions, frikelQuinto2015_limitedViewTomoPAT} in PAT with finite omnidirectional flat sensor ($\theta_\text{max} = \pi/2$). The limitations of CS approaches with Curvelet sparsity regularisation were discussed in \cite{frikel2010_limitedXray, frikel2013sparse} in the context of X-ray tomography.

Motivated by these works, a number of authors considered implications of the visible and invisible singularities for learning methods in limited-angle tomography in X-ray and in general \cite{bubba2019learning,bubba2021deep,andrade2021deep}. 
The approach proposed in  \cite{andrade2021deep} capitalizes on the microlocal properties of the Radon transform 
to inpaint the singularities extracted from the data, using network proposed in \cite{DENSE}, into the reconstructed image.
In \cite{bubba2021deep} the authors reinterpret the unrolled ISTA iteration as CNN layers which coefficients and structure are inferred from the convolutional properties of Fourier integral operators and pseudo-differential operators and apply it to limited-angle X-ray tomography. The present work is closest in spirit to the approach proposed in \cite{bubba2019learning} for the limited-angle X-ray tomography which promotes the idea to separate the visible part of the image, which can be stably reconstructed in Shearlets frame using CS approach, from the invisible part of the image which needs to be learned. The main benefits of such approach are i) a clear separation between what is stably reconstructed from the measured data and the prior that is learned from the training set and ii) decoupling of the reconstruction of the visible which requires iterating with the forward/adjoint operator from learning the invisible which does not, as the invisible is contained in the null space of the discrete forward operator. A similar idea was proposed from an algebraic perspective as null space learning in \cite{schwab2019deep}. The directional frames like Curvelets or Shearlets have the significant advantage of providing a microlocal representation of the null space and its complement.

\subsection{Motivation}
\label{sec:sec:motivation}
\dontshow{In PAT with flat sensor, there are two types of limited-view problems: i) due to finite size of the sensor 
and ii) due to varying sensitivity of the sensor to the direction of the wavefront impinging on the detector. In this work we focus on the second type.
}
In this work we focus on the limited-view problem due to varying sensitivity of the sensor to the direction of the wavefront impinging on the detector.
The directional sensitivity can be modelled as binary with a maximal impingement angle $\pm\theta_{\max}, \, \theta_{\max} < \pi/2$, for the wavefronts to be registered, resulting in a cone of sensitivity around the direction normal to the sensor. We observe this problem shares many similarities with the limited-angle parallel X-ray, in particular the decomposition into visible/invisible singularities in the Fourier domain corresponds to the partition of the angular range of the singularities regardless of their location, exactly as for the parallel X-ray transform. We note that this is not the case for PAT with a flat sensor of finite size where angular range of visible singularities varies with the position of the singularity.

The decomposition into visible and invisible singularities for limited-angle parallel X-ray geometry underpins the result in \cite{frikel2013sparse} on visible singularities only recovery via compressed sensing in Curvelets or Shearlets frames.
Based on this result, the  authors of \cite{bubba2019learning} proposed to reconstruct the visible and learn the invisible in Shearlet frame for the limited-angle parallel X-ray tomography. These results along with the above observation of similarities between the limited-angle parallel X-ray CT and PAT with flat sensor with limited sensitivity angle, \emph{limited-angle PAT} for short, 
motivate our work. 

\subsection{Contribution}
\label{sec:sec:contribution}
Using the wavefront mapping \cref{eq:WaveFrontMapping} \cite{pan2021photoacoustic}, we obtain the correspondence between the visible and invisible wavefront directions in PAT image and PAT data. 
This allows us the observation that the limited sensitivity angle restricts the data range in a way that enables the use of efficient Fourier domain forward/adjoint and pseudoinverse operators derived in \Cref{sec:sec:FourierOperators,appendixlabel1,sec:sec:LimitedAngleFourierOperators}.

\dontshow{
Due to the angular range visible/invisible partition property of the limited-angle PAT we can adapt the wedge restricted Curvelet transform, proposed in  \cite{pan2021photoacoustic} for PAT data representation, to the representation of the visible/invisible components of the PAT image (initial pressure, $p_0$), and recover the visible singularities via CS 
in a wedge restricted Curvelet frame with the data fidelity term using the aforementioned limited-angle PAT Fourier operators. The visible component of the PAT image is then
represented by its Coronae coefficients, a minimal representation which perfectly matches the split into the visible/invisible and the multiscale structure induced by the Curvelet frame, introduced in \Cref{sec:sec:sec:CoronaeInput}. These visible Coronae coefficients are
fed into a Coronae-Net, a tailored U-Net inspired architecture, designed and trained to fill in the invisible component of the PAT image given the visible component. 
}

Due to the visible/invisible inducing an angular range partition in the limited-angle PAT, we can adapt the wedge restricted Curvelet transform proposed in \cite{pan2021photoacoustic} for a sparse PAT data representation, to a sparse representation of the visible/invisible components of the PAT image (see \Cref{sec:sec:FWRCurvelets}) and recover the sparse wedge restricted Curvelet coefficients of the visible image component 
via CS 
with the data fidelity term using the aforementioned limited-angle PAT Fourier operators. 
The recovered visible component of the PAT image is then
represented by its Coronae coefficients, visible Coronae coefficients for short (see \Cref{sec:sec:sec:CoronaeDR,sec:CoronaeVsCurvelets}), resulting in a minimal representation which perfectly matches the split into the visible/invisible and inherits the multiscale structure of the Curvelet frame.
These visible Coronae coefficients are
fed into a Coronae-Net, a tailored U-Net inspired architecture, designed and trained to fill in the invisible component of the PAT image given the visible component. 

\subsection{Outline}
\label{sec:sec:outline}
The reminder of the paper is organized as follows. \Cref{sec:PAT} briefly introduces the time domain formulation of the forward, adjoint and inverse problems in PAT. In \Cref{sec:PATFourierOperators} we summarise the existing results on Fourier domain PAT and stipulate the limited-angle Fourier PAT operators. The new, to best of our knowledge, derivation of the adjoint in Fourier domain is presented in \cref{appendixlabel1}. In \Cref{sec:CurveletRepresentationImage} we introduce the various multiscale representations used throughout the paper, discuss their roles and mutual relations and justify their suitability for representation of the initial pressure in limited-angle PAT. In \Cref{sec:ReconVisLearnInv} we outline our framework for reconstructing the visible via the solution of a CS problem using the Fourier domain limited-angle PAT operators and a sparse representation in a fully wedge restricted Curvelet frame, and learning the invisible via the proposed Cornae-Net working on Coronae coefficients of the visible and invisible parts of the initial pressure image. We also extend the framework to real world scenario with imperfect visible coefficients via residual learning,  ResCornae-Net. We illustrate the performance of our framework in both perfect and imperfect learning scenarios and its generalisation potential on a fully synthetic ellipse data set in \Cref{sec:simulatedexperiments} and on a more realistic DRIVE retina vessel data set in \Cref{sec:RealDataApplication}. \Cref{sec:Conclusion} provides conclusions and outlook.

\section{Photoacoustic Tomography}
\label{sec:PAT}
With several assumptions on the tissue properties \cite{wang2011photoacoustic}, the photoacoustic forward problem can be modelled as an initial value problem for the free space wave equation
\begin{equation}
\label{eq:PATforwardequation}\tag{{\bf A}}
\begin{aligned}
\nonumber \underbrace{\left( \frac{1}{c^2(\bx)}\frac{\partial^2}{\partial t^2}  - \nabla^2 \right)}_{:=\square^2 } p(t,\bx) &= 0, \quad (t,\bx)\in (0,T)\times\mathbb{R}^d, \\
p(0,\bx) & = p_0(\bx), \\
p_t(0,\bx) &  = 0,
\end{aligned}
\end{equation}
where $p(t,\bx) \in \mathcal{C}^\infty((0,T)\times\mathbb{R}^d)$ is a time dependent acoustic pressure recorded for a finite time $T$, $p_0(\bx) \in \mathcal{C}_0^\infty(\mathbb{R}^d)$ is its initial value, and $c(\bx) \in \mathcal{C}^\infty(\mathbb{R}^d)$ is the speed of sound in the tissue.

The photoacoustic inverse problem recovers the initial pressure $p_0(\bx)$ 
in the region of interest $\Omega$ on which $p_0(\bx)$ is compactly supported, from time dependent measurements 
\begin{equation*}
g(t,\bx_{\mathcal{S}}) = \omega(t) p(t,\bx_\mathcal{S}), \quad (t,\bx_{\mathcal{S}})\in (0,T)\times \mathcal S
\end{equation*} 
at a set of locations $\bx_{\mathcal S} \in \mathcal{S} \subset \mathbb{R}^d$, e.g.~the boundary of $\Omega$, where $\omega\in \mathcal{C}_0^{\infty}(0,T)$ is a temporal smooth cut-off function. It amounts to a solution of the following initial value problem for the wave equation with constraints on the surface \cite{finch2004determining}
\begin{equation}\label{eq:TRequation}\tag{\bf TR}
\begin{aligned}
\square^2 q(t,\bx) &=0, & \text{ }  (t,\bx)\in (0,T)\times\mathbb R^d, \\
q(0,\bx) & = 0, \; q_t(0,\bx) = 0, & \text{ }  \\
q(t,\bx) & = g(T-t,\bx_{\mathcal{S}}), & \text{ }(t,\bx_{\mathcal{S}})\in (0,T)\times\mathcal S,
\end{aligned}
\end{equation}
evaluated at $T$, $q(T,\bx)$, also referred to as a \emph{time reversal}. For non-trapping smooth $c(\bx)$, in 3D, the solution of \cref{eq:TRequation} is the solution of the PAT inverse problems if $T$ is chosen large enough so that $g(t,\bx) = 0$ for all $\bx\in\Omega$, $t\geq T$ and the wave has left the domain $\Omega$. Assuming that the measurement surface $\mathcal S$ surrounds the region of interest $\Omega$ containing the support of initial pressure $p_0$, the wave equation \cref{eq:TRequation} has a unique solution.

When no complete data is available (meaning that some singularities have not been observed at all as opposed to  the case where partial energy has been recorded, see e.g.~\cite{stefanov2009thermoacoustic}), $p_0$ is usually recovered in a variational framework, including some additional penalty functional corresponding to prior knowledge about the solution. Iterative methods for solution of such problems require application of the adjoint operator which amounts to the solution of the following initial value problem with a time varying mass source \cite{arridge2016adjoint}
\begin{equation}\label{eq:adjoint}\tag{\text{$\textbf{A}^\star$}}
\begin{aligned}
\square^2 q(t,\bx) &= \left\{
\begin{array}{cc} 
\frac{\partial}{\partial t} g(T-t,\bx_{\mathcal{S}}), &(t,\bx_{\mathcal{S}})\in  (0,T)\times\mathcal S\\ 
0 & \text{everywhere else} 
\end{array}
\right.\\
q(0,\bx) & = 0, \\
q_t(0,\bx) & = 0,
\end{aligned}
\end{equation}
evaluated at $T$, $q(T,\bx)$.
\dontshow{
All the equations \cref{eq:PATforwardequation}, \cref{eq:adjoint} and \cref{eq:TRequation} can be solved using the pseudospectral method implemented in the k-Wave toolbox for heterogeneous media in two and three dimensions \cite{treeby2010k}; see \cite{arridge2016adjoint} for implementation of the adjoint \cref{eq:adjoint} 
in this setting.}

\section{Fourier Domain Formulation of PAT}
\label{sec:PATFourierOperators}
The operators defined in this section map between Fourier transforms of the initial pressure $p_0$ and the data $g$. 
When using Fourier domain PAT operators in conjunction with Curvelet frame (which is also defined and computed via the Fourier transform), all the computations can be carried out directly in the Fourier domain without need to compute intermediate Fourier transforms.


\subsection{Forward, Adjoint and Inverse PAT Operators in the Fourier Domain}
\label{sec:sec:FourierOperators}
We consider a PAT geometry with a line sensor in 2D (a plane sensor in 3D). Let $\bx = (\bx_\perp, \bx_{\mathcal S}) \in \mathbb R \times \mathbb R^{d-1}$ and the sensor hyperplane be $\mathcal{S} = \{\bx \in \mathbb R^{d}:  \bx_\perp = 0 \}$. Then  
$\bx_{\mathcal S}$ describes the component parallel to the sensor and $\bx_\perp$ the component perpendicular to the sensor. 
Assuming half space symmetry with respect to the sensor plane
with a constant speed of sound $c(\mathbf{x}) = c$, the solution of the wave equation
\cref{eq:PATforwardequation} in the Fourier domain can be explicitly written as \cite{cox2005fast}
\begin{equation}\label{eq:p0cos}
\hat p(ct,\bk) = \cos(c|\mathbf{k}|t)\hat  p_0(\bk),
\end{equation}
where $\hat{u}$ denotes the Fourier transform of $u$ in some or all variables and $\mathbf{k} \in \mathbb R^d$ is the Fourier domain wave vector. Using \cref{eq:p0cos}, we can obtain the following \emph{forward} mapping between the initial pressure $p_0(\bx_\perp, \bx_{\mathcal S}), \, (\bx_\perp, \bx_{\mathcal S}) \in \mathbb R \times \mathbb R^{d-1}$ and the PAT data $g(ct, \mathbf{x}_S), \, (ct, \mathbf{x}_S) \in \mathbb R \times \mathbb R^{d-1}$, with $g(ct, \mathbf{x}_S) = p(ct, \mathbf{x}_S)$\footnote{Note, the constant speed of sound $c$ is used to normalise the time/frequency in our Fourier formulation of the operators, the non-normalised (physical) formulation of the forward operator would have an extra factor $1/c$ on the right hand side of \ref{eq:FrequencyEquation} (s.a.~\cite{pan2021photoacoustic}) which in turn would affect the scaling of the other operators. It has the consequence that $g$, as defined here, needs to be scaled by $1/c$ to correspond to physical PAT data which is what we do in simulations.}, in the Fourier domain  \cite{cox2005fast,kostli2001temporal,xu2002exact,Acoustic}
\begin{align}
\label{eq:FrequencyEquation}\tag{\text{$\hat{\textbf{A}}$}} \hat g(\omega/c, \bk_{\mathcal{S}}) &= \frac{\omega/c}{\sqrt{(\omega/c)^2-{|\mathbf{k}_{\mathcal{S}}}|^2}} \hat p_0\left(\sqrt{ (\omega/c)^2- |\mathbf{k}_{\mathcal{S}}}|^2, \bk_{\mathcal{S}}\right),
\end{align}
where 
$\bk = (\bk_\perp, \bk_{\mathcal S})$ are the Fourier domain  counterparts to the ambient space variables $\bx = (\bx_\perp, \bx_{\mathcal S})$.
In the above formulae the time $t$, its frequency $\omega$ and $\bk_\perp$ are understood as non-negative quantities, with symmetries defining the functions on their negatives. In particular, formula \cref{eq:p0cos} has an implicit assumption of even symmetry of $p_0(\bx_\perp, \bx_S ), \, g( ct, \bx_S)$ in $\mathbb R^d$ w.r.t.~$\bx_\perp = 0, \, ct = 0$, respectively, 
\begin{align}\label{eq:map:k:mir}
\hat p_0 (-\bk_\perp, \bk_S)  &= \hat p_0  ( \bk_\perp, \bk_S), \\
\nonumber \hat g  (-\omega/c, \bk_S) &= \hat g  (\omega/c, \bk_S ).
\end{align}
For the right hand side of equation \cref{eq:FrequencyEquation} to be well-defined and bounded, the expression under the square root should be bounded away from zero for all non-zero wave vectors i.e.~$\mathcal R_\textbf{A} = \left\{ (\omega/c, \bk_{\mathcal S})  \in \mathbb R^d: |\mathbf{k}_{\mathcal{S}}|< |\omega/c |\right\} \cup {\bf 0} $, resulting in a bow-tie shaped range of photoacoustic operator in the sound-speed normalised Fourier space, see \cref{image:BowtieTheta60} (b). %
\dontshow{which is illustrated by the contour plot of $\bk_\perp=\mbox{sign}(\omega)\sqrt{(\omega/c)^2-|\mathbf{k}_{\mathcal{S}}|^2}$, $\bk_\perp = const$, in \cref{image:Bowtie}.} We note that \cref{eq:FrequencyEquation} turned around underlies the inverse transform (which is stable) \cite{kostli2001temporal}, but as forward transform the factor blows up and regularisation is required \cite{hauptmann2018approximate}.

\dontshow{
\begin{figure}[htbp!]
  \centering
  \includegraphics[width=0.48\linewidth,height=0.34\linewidth]{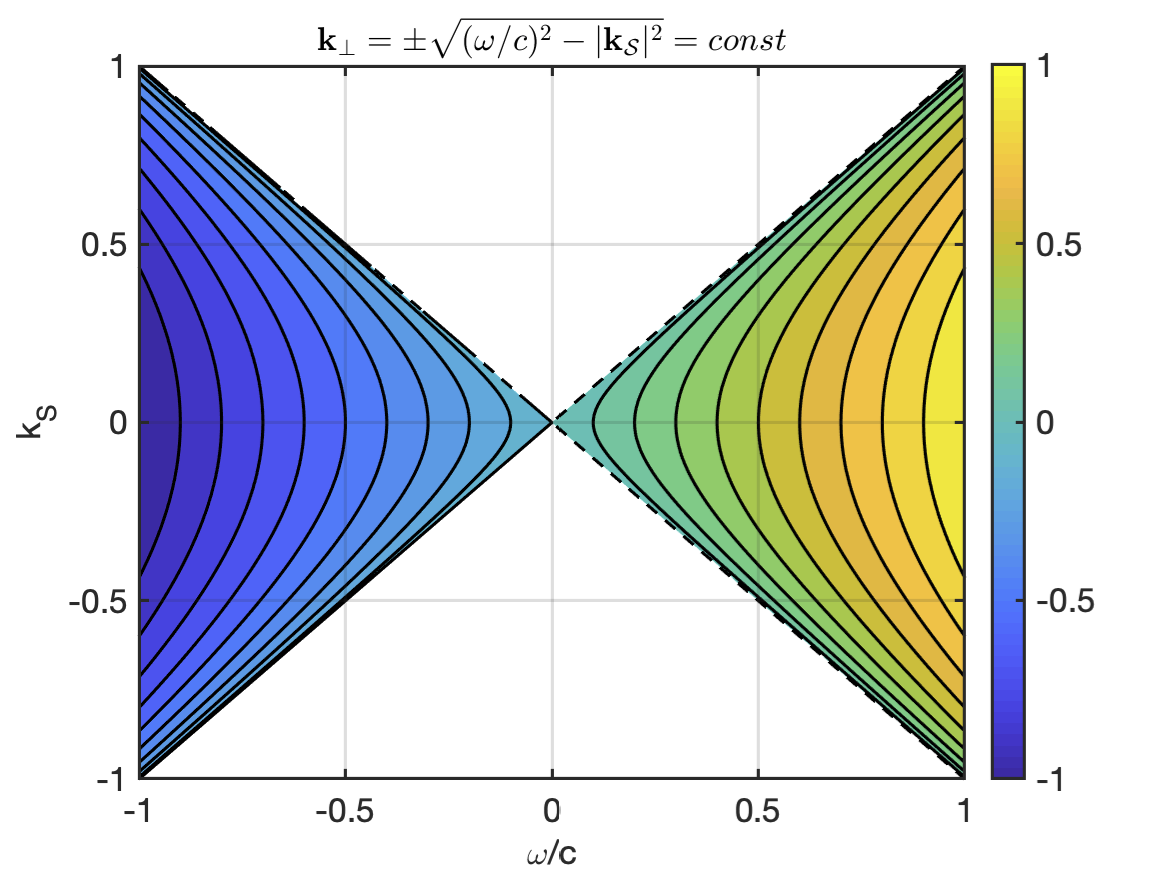}
  \caption[image:Bowtie]{Contour plot of $\bk_\perp=\mbox{sign}(\omega)\sqrt{(\omega/c)^2-|\mathbf{k}_{\mathcal{S}}|^2} = const$ over the bow-tie shaped frequency domain range of the photoacoustic Fourier forward operator, 
  $\mathcal R_A = \left\{ (\omega/c, \bk_{\mathcal S}) \in \mathbb R^d: |\mathbf{k}_{\mathcal{S}}|<\omega/c\right\} \cup {\bf 0}$.} 
  \label{image:Bowtie}
\end{figure}
}

As already alluded, the inverse PAT Fourier operator follows immediately from \cref{eq:FrequencyEquation}, 
\begin{equation}\label{eq:FrequencyEquationInv}\tag{\textbf{$\hat{\textbf{A}}^{-1}$}}
\hat p_0(\bk_{\perp}, \bk_{\mathcal{S}}) = \frac{|\bk_{\perp}|}{|\bk|} \hat g\left(|\bk|,\bk_\mathbf{S}\right),
\end{equation}
with the
change of variables induced by the dispersion relation in the wave equation $(\omega/c)^2 = |\bk|^2 = |\bk_{\mathcal S}|^2 + |\bk_{\perp}|^2$,
\begin{align}\label{eq:map:k}
\bk = (\bk_\perp, \bk_S ) &\leftarrow \left( \sign{(\omega)} \sqrt {(\omega/c)^2 - |\bk_S|^2 }, \bk_S \right), \\
\nonumber (\omega/c, \bk_S) &\leftarrow \left( \sign{(\bk_\perp)}  \left(|\bk| =   \sqrt { |\bk_\perp|^2 + |\bk_S|^2 } \right), \bk_S \right).
\end{align}
The even symmetry \cref{eq:map:k:mir} is inherited by the map \cref{eq:map:k}.

Using the definition of the adjoint, the adjoint PAT Fourier operator was derived in the \cref{appendixlabel1}
\begin{equation}\label{eq:FrequencyEquationAdj}\tag{\text{$\hat{\textbf{A}}^*$}}
\hat p_0(\bk_{\perp}, \bk_{\mathcal{S}}) = \hat g\left(|\bk|,\bk_\mathbf{S}\right).
\end{equation}
We note that both inverse \cref{eq:FrequencyEquationInv} and adjoint \cref{eq:FrequencyEquationAdj} operators only differ by the scaling factor present in the inverse and absent from the adjoint, while the underlying change of variables is the same.

\subsection{Wavefront Direction Mapping}
\label{sec:sec:WfM}
\begin{figure}[htbp!]
  \centering
  \subfloat[]{
  \includegraphics[width=0.5\linewidth,height=0.28\linewidth]{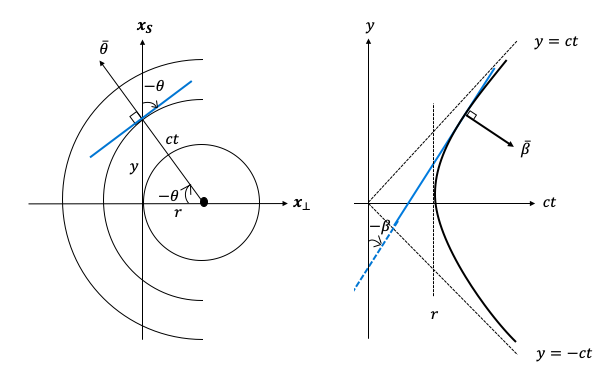}}
  \subfloat[]{
  \includegraphics[width=0.48\linewidth,height=0.27\linewidth,trim={0 {0.03\linewidth} 0 0}, clip=true]
  {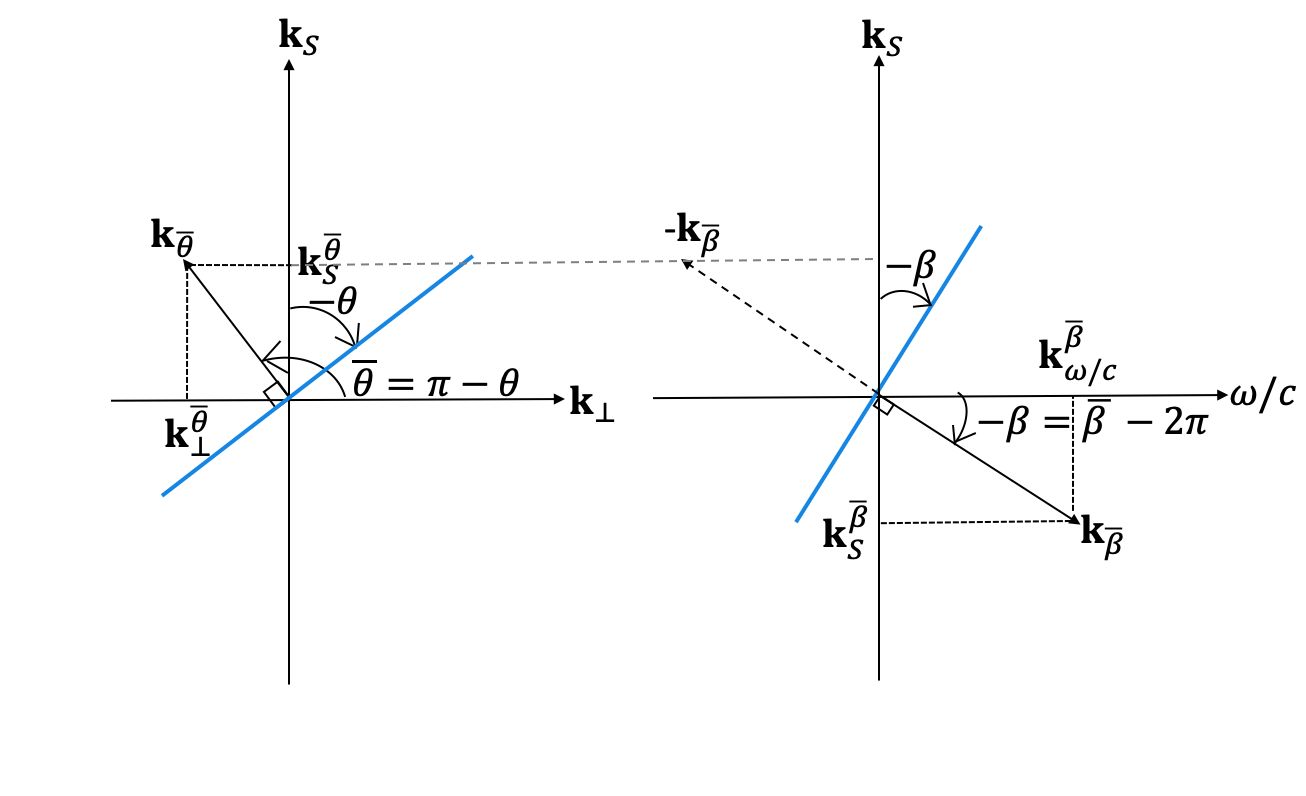}}
  \caption{(a) Wavefront direction mapping between the ambient space $\bar\theta$ (left), and the data space, $\bar\beta$ (right) on an example of a 2D spherical wave. 
  $-\theta$ and $-\beta$ are the angles that the wavefronts (in \textcolor{cyan}{blue}) make with the sensor (the negative signs are the consequence of the right handed coordinate system). 
  (b) The Fourier domain counterpart, the wavefront vector mapping between $\bk_{\bar\theta}$ (left), and $\bk_{\bar\beta}$ (right).}
  \label{image:PATforwardmodel}
\end{figure}

The wavefront mapping \cref{eq:WaveFrontMapping} was derived in \cite{pan2021photoacoustic} observing that the change of coordinates \cref{eq:map:k}
defines a one-to-one map between the frequency vectors $\bk = (\bk_\perp, \bk_S)  \in \mathbb R^d$ and $( \omega/c, \bk_S) \in \mathcal R_\textbf{A}$.
\dontshow{
\begin{align}\label{eq:map:k}
\bk = (\bk_\perp, \bk_S ) &\leftarrow \left( \sqrt {(\omega/c)^2 - |\bk_S|^2 }, \bk_S \right), \\
\nonumber (\omega/c, \bk_S) &\leftarrow \left(|\bk| = \sqrt { |\bk_\perp|^2 + |\bk_S|^2 }, \bk_S \right).
\end{align}
}
This mapping is illustrated in \cref{image:PATforwardmodel}.  \cref{image:PATforwardmodel} (a) shows the wavefronts (in \textcolor{cyan}{blue}) and their corresponding wavefront vectors in the ambient space $\bar{\theta} = (-\cos\theta, \sin\theta)$  (on the left) and in the data space $\bar{\beta} = (\cos\beta, -\sin\beta)$ (on the right). Note, that these wavefront vectors have the same direction (and hence angles\footnote{For simplicity of notation, we do not distinguish between the ambient/data space vectors $\bar\theta, \bar\beta$ and their angles in the right handed coordinate system. If the vector or its angle is meant depends on the context.}: $\bar{\theta}$, $\bar{\beta}$) as their frequency domain counterparts $\bk_{\bar\theta}$, $ \bk_{\bar\beta}$ which are depicted in \cref{image:PATforwardmodel} (b).
\dontshow{
\begin{align}\label{eq:map:k:mir}
\hat p_0 (-\bk_\perp, \bk_S)  &= \hat p_0  ( \bk_\perp, \bk_S), \\
\nonumber \hat p  (-\omega/c, \bk_S) &= \hat p  (\omega/c, \bk_S ).
\end{align}
}
\dontshow{
\begin{figure}[htbp!]
  \centering
  \includegraphics[width=0.8\linewidth,height=0.3\linewidth]{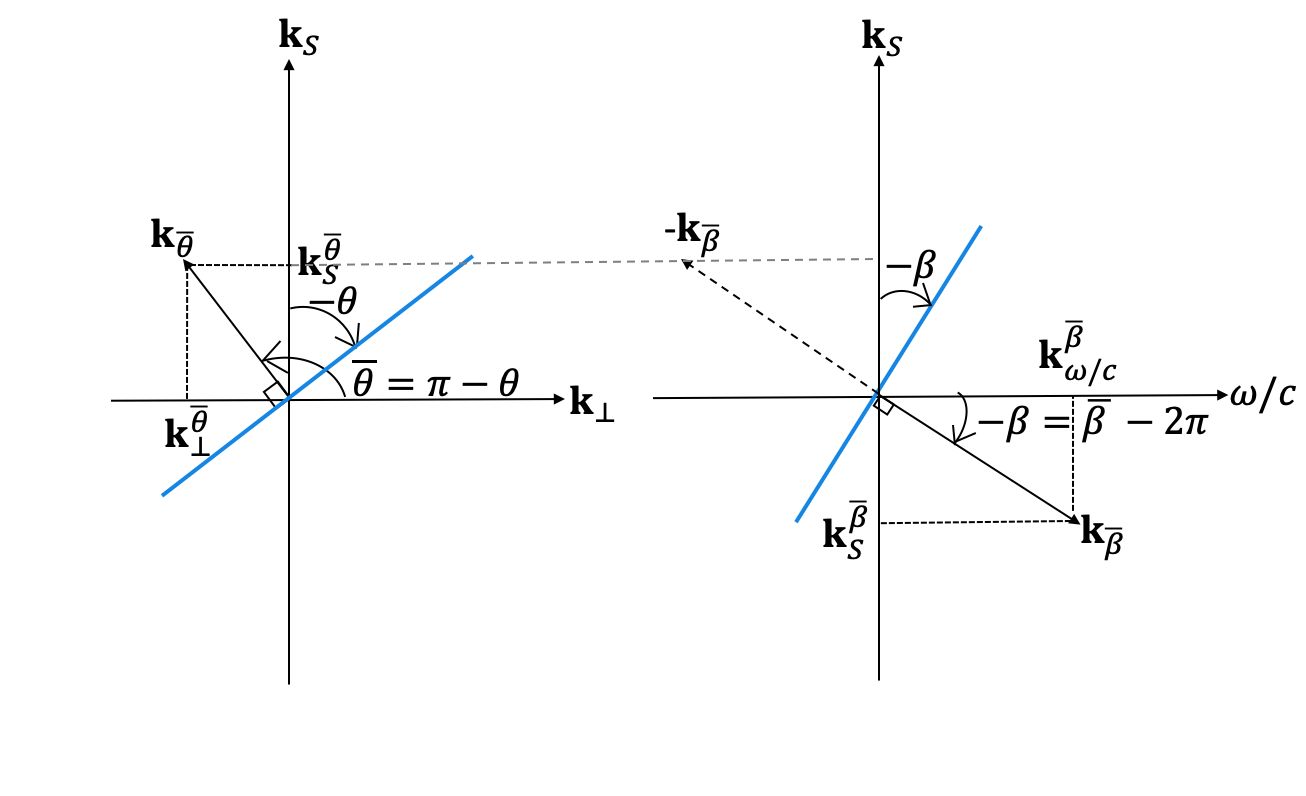}
  \caption{Frequency domain visualisation of the example in \cref{image:PATforwardmodel}. Fourier domain wavefront vector mapping between the ambient space, $\bk_{\bar\theta}$ (left), and the data space, $\bk_{\bar\beta}$ (right).}
  \label{image:PATforwardmodel:k}
\end{figure}}
With the notation in \cref{image:PATforwardmodel}, 
we deduce the map between the wavefront vector angles $\bar{\theta}$ and $\bar{\beta}$ considering the corresponding frequency domain vectors
\begin{align}\label{eq:map:kangles}
\tan \bar\beta &= \frac{\bk_{\mathcal S}^{\bar\beta}}{\bk_{\omega/c}^{\bar\beta}}= \frac{-\bk_{\mathcal S}^{\bar\theta}}{|\bk_{\bar\theta}| =\sqrt{(|\bk_\perp^{\bar\theta}|^2 + |\bk_{\mathcal S}^{\bar\theta}|^2 }} = \sin(-\bar\theta).
\end{align}
We can express the wavefront mapping \cref{eq:map:kangles} in terms of the angles the wavefronts make with the detector $-\theta$, $-\beta$, using their relation with wavefront vector angles $\bar\theta = \pi -\theta$, $\bar\beta = 2\pi -\beta$ and basic trigonometric identities
\begin{equation}\label{eq:WaveFrontMapping}\tag{\textbf{WfM}}
\beta = \arctan\left(\sin\theta \right),
\end{equation}
where $\theta \in (-\pi/2,\pi/2)$ and $\beta \in \left(-\pi/4, \pi/4\right)$ for the un-mirrored initial pressure. The derivation was presented in $\mathbb R^d, \, d=2$ for simplicity, $d=3$ follows analogously by treating both detector coordinates the same, see also \cite{pan2021photoacoustic}.

\subsection{Limited-angle PAT Fourier Operators}
\label{sec:sec:LimitedAngleFourierOperators}
According to the wavefront mapping \cref{eq:WaveFrontMapping} we have 
$$\tan \beta_\text{max} = \sin \theta_\text{max},$$ 
which yields the correspondence between the visible ambient space wavefronts $[-\theta_\text{max},\theta_\text{max}]$,\\ $0 \leq  \theta_\text{max}<\pi/2$ and the visible data space wavefronts $\beta\in [-\beta_\text{max},\beta_\text{max}]$,$0 \leq \beta_\text{max}<\pi/4$. Thus we only need to
restrict the range of $\theta$ 
in equation \cref{eq:FrequencyEquation} to formulate the limited-angle PAT Fourier forward operator 
\begin{equation}\label{eq:FrequencyEquationReg}\tag{\text{$\hat{\textbf{A}}_\angle$}}
\begin{aligned}
\hat g(\omega/c, \bk_{S}) & = \frac{\omega/c}{\sqrt{(\omega/c)^2-{|\mathbf{k}_{S}}|^2}} \hat p_0\left(\sqrt{ (\omega/c)^2- |\mathbf{k}_{S}}|^2, \bk_{S}\right), \quad |\bk_S| \leq \sin(\theta_{\max}) |\bk|.
\end{aligned}
\end{equation}
\cref{eq:FrequencyEquationReg} maps from the domain $\mathcal D_\textbf{A}^\angle = \left\{ (\bk_{\perp}, \bk_S)  \in \mathbb R^d: |\bk_S| \leq \sin(\theta_{\max}) |\bk| \right\}$, 
$\sin \theta_{\max} < 1$
onto the narrowed bow-tie shaped range 
$\mathcal R_\textbf{A}^\angle = \left\{ (\omega/c, \bk_S)  \in \mathbb R^d: |\mathbf{k}_S| \leq |\omega/c |\tan \beta_\text{max} \right\}$, 
$\tan \beta_\text{max} \\ <1$, where again we extend the map by even symmetry for $\omega/c < 0$.

\begin{figure}[htbp!]
  \centering
  \subfloat[]{
  \includegraphics[width=0.4\linewidth,height=0.3\linewidth]{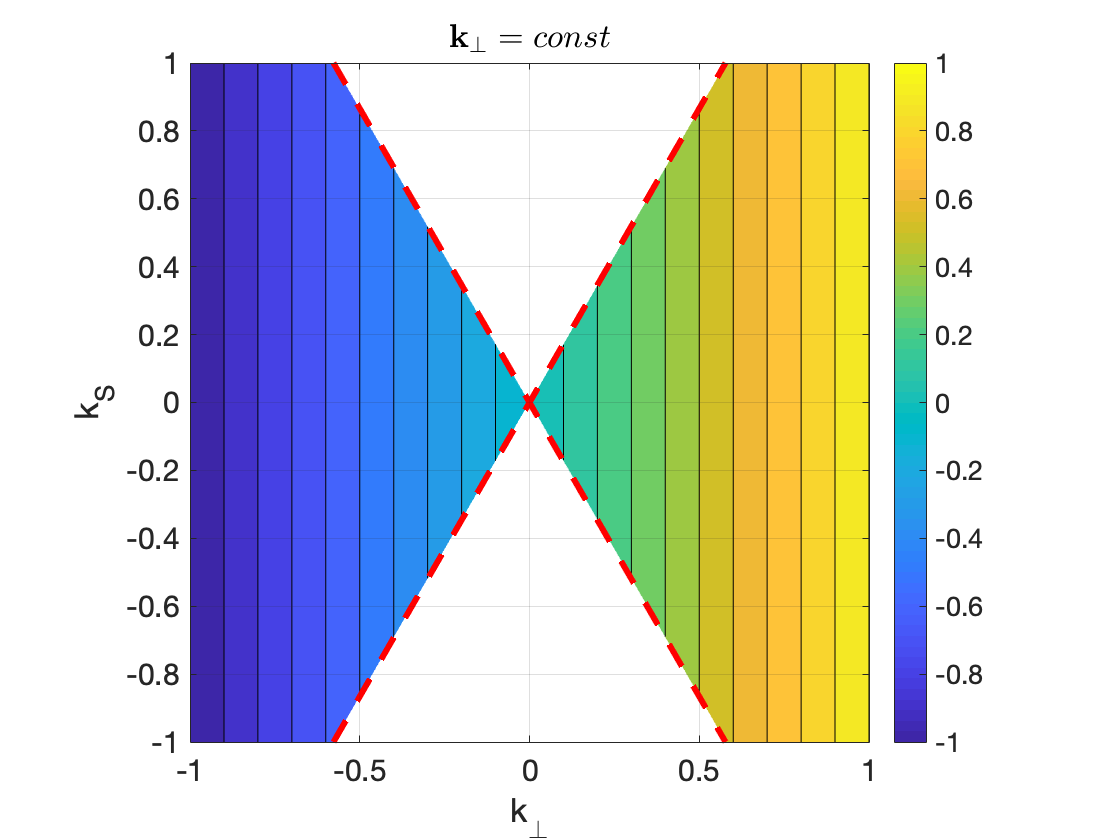}}
  \subfloat[]{
  \includegraphics[width=0.4\linewidth,height=0.3\linewidth]{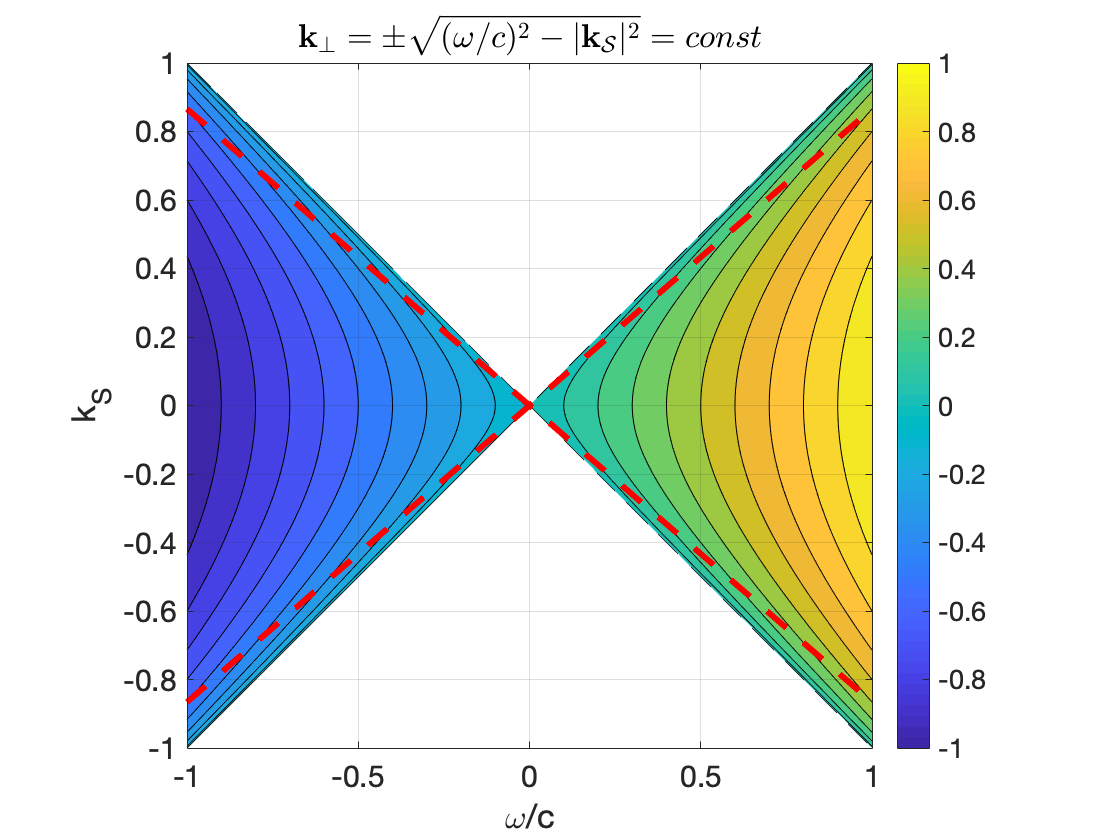}}
  \caption[image:bowtieTheta60]{Contour plot of $\bk_\perp = const$ over the bow-tie shaped (a) domain of the limited-angle PAT Fourier forward operator, $\theta_{\max} = \pi/3$,
  (b) range of the PAT Fourier forward operator with the further narrowed range of the limited-angle PAT Fourier forward operator highlighted with the \textcolor{red}{red dashed} line.
  } 
  \label{image:BowtieTheta60}
\end{figure}

A Fourier domain illustration of the visibility in ambient space for $\theta_\text{max} = \pi/3$ is shown in \cref{image:BowtieTheta60} (a), and the corresponding narrowing down of the bow-tie shaped range in the Fourier transformed data space is highlighted with the red cut-off line in \cref{image:BowtieTheta60} (b).
With this restriction $|\theta| \leq \theta_{\max} < \pi/2$, the denominator of \cref{eq:FrequencyEquationReg} is bounded away from 0,
\begin{equation}\label{eq:singularity}
(\omega/c)^2 - |\bk_{\mathbf{S}}|^2 \geq (1-\tan^2\beta_{\text{max}})(\omega/c)^2 = (\omega/c)^2\cos^2\theta_{\text{max}}
>0,
\end{equation}
and the limited-angle PAT Fourier forward operator can be stably evaluated. We note, that our restriction is equivalent to an upper bound on the factor (suggested as regularisation in \cite{hauptmann2018approximate})
\begin{equation}\label{eq:WeightingFactor}
\frac{\omega/c}{\sqrt{(\omega/c)^2-{|\mathbf{k}_{S}}|^2}} = \frac{|\bk|}{|\bk_\perp|} = \frac{1}{\cos\theta} \leq D,
\end{equation}
with the choice of $D = (\cos\theta_\text{max})^{-1}$.

We can now analogously define the limited-angle PAT Fourier inverse operator, $\mathcal R_\textbf{A}^\angle \to \mathcal D_\textbf{A}^\angle$
\begin{equation}\label{eq:FrequencyEquationInvReg}\tag{\text{$\hat{\textbf{A}}^{-1}_\angle$}}
\hat p_0(\bk_{\perp}, \bk_{S}) = \frac{|\bk_{\perp}|}{|\bk|} \hat g\left(|\bk|,\bk_\mathbf{S}\right), \quad |\bk_S| \leq \sin(\theta_{\max}) |\bk|, 
\end{equation}
and the limited-angle PAT Fourier adjoint operator, $\mathcal R_\textbf{A}^\angle \to \mathcal D_\textbf{A}^\angle$
\begin{equation}\label{eq:FrequencyEquationAdjReg}\tag{\text{$\hat{\textbf{A}}^*_\angle$}}
\hat p_0(\bk_{\perp}, \bk_{S}) = \hat g\left(|\bk|,\bk_\mathbf{S}\right), \quad |\bk_S| \leq \sin(\theta_{\max}) |\bk|. 
\end{equation}
We note, that by Schwartz's Paley-Wiener theorem a Fourier transform of a compactly supported function is an entire function, thus in principle the function can be recovered from the knowledge of its Fourier transform on any open set, a.k.a.~the continuous limited-angle operator is injective on compactly supported functions. However, a numerical realisation of such procedure would be highly unstable.
In practice the functions $p_0$, $g$ are discretized on a grid in $\mathbb R^d$ and a discrete fast Fourier transforms along with interpolation are used for evaluation of the operators. Then the discrete limited-angle forward operator has a non-trivial null space which corresponds to the invisible part of the image. In what follows, we use the same notation for the continuous and discrete operators, as their meaning is clear from the context.

\begin{figure}[htbp!]
\centering
  \subfloat[$p_0$]{
  \includegraphics[width=0.21\linewidth,height=0.18\linewidth]{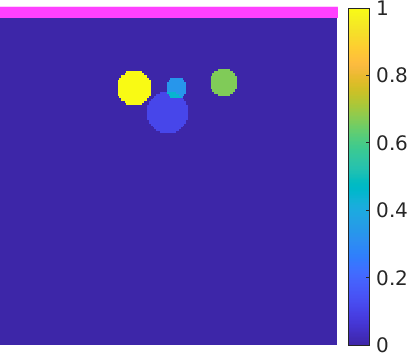}}
  \subfloat[$g_\rightangle$]{
  \includegraphics[width=0.21\linewidth,height=0.19\linewidth]{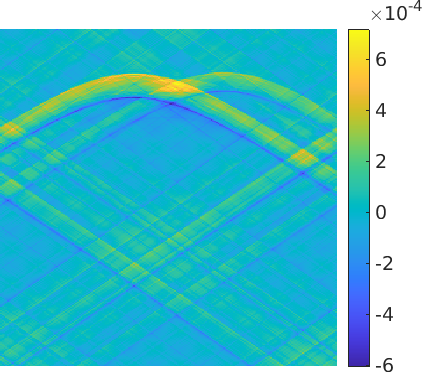}}
  \subfloat[$g_\angle$]{
  \includegraphics[width=0.21\linewidth,height=0.19\linewidth]{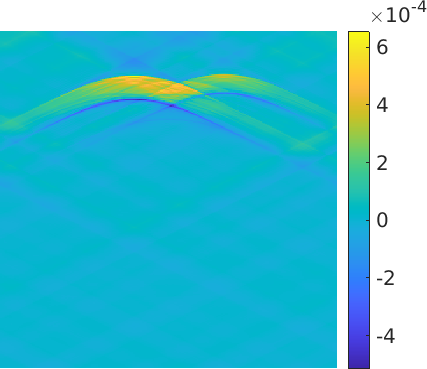}}
  \\
  \subfloat[${p_0}^\text{Linear}$]{
  \includegraphics[width=0.21\linewidth,height=0.18\linewidth]{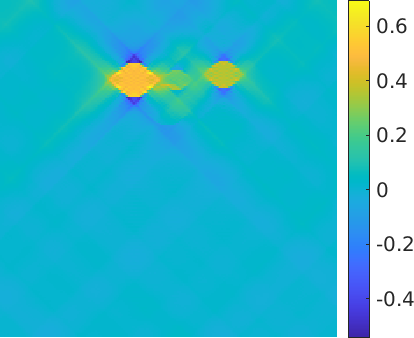}}
  \subfloat[${p_0}^\text{Adj}$]{
  \includegraphics[width=0.21\linewidth,height=0.18\linewidth]{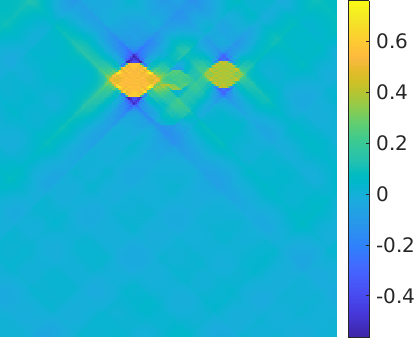}}
\subfloat[$p_0^\text{Linear} - p_0^\text{Adj}$]{
  \includegraphics[width=0.21\linewidth,height=0.18\linewidth]{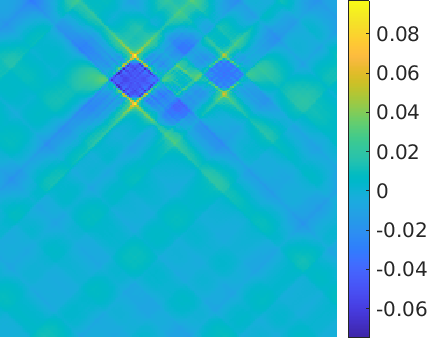}}
  \caption[BallsFB]{4-disk Phantom - Limited-angle PAT Fourier operators: (a) $p_0$: 4 disks in the north sector of a square domain with a line detector on top (in \textcolor{magenta}{pink}), (b) full-angle data $g_\rightangle$ obtained via PAT Fourier forward operator $g_\rightangle = \hat{\textbf{A}}p_0$ (a.k.a.~$\theta_\text{max} \approx \pi/2$) and (c) limited-angle data $g_\angle =  \hat{\textbf{A}}_{\angle}p_0$ with $\theta_\text{max} = \pi/4$, (d) direct linear inversion of limited-angle data, $g_\angle$, $p_0^\text{Linear} = \hat{\textbf{A}}^{-1}_\angle g_{\angle}$, (e) backprojection of $g_\angle$, $p_0^\text{Adj} = \hat{\textbf{A}}^*_\angle g_\angle$ and (f) difference between the direct linear inversion and the backprojection of $g_\angle$, $p_0^\text{Linear} - p_0^\text{Adj}$.

  \dontshow{, and (f) Neumann series reconstruction of $g_\angle$ after 20 iterations.}}
  \label{image:BallsFB}
\end{figure}

\dontshow{Finally, we consider a non-variational approach based on the Neumann series \cite{stefanov2009thermoacoustic, qian2011efficient} which is known to effectively restore partially visible singularities which arise in limited-view problems. The $k$-th iterate in the Neumann series approach with the limited-angle PAT Fourier operators is obtained as 
\begin{equation}\label{eq:Neumann}\tag{\textbf{iN}}
p^k_\textbf{iN} = \sum_{i=0}^k K^i \hat{\textbf{A}}^{-1}_\angle g_\angle, \quad \text{with} \quad K = I - \hat{\textbf{A}}^{-1}_\angle \hat{\textbf{A}}_\angle,
\end{equation}
where $I$ is the identity operator over $\mathbb{R}^d$ and $g_\angle = \hat{\textbf{A}}_\angle p_0$ is the limited-angle data (here noiseless). The reformulation of \cref{eq:Neumann} leads to an iterative image reconstruction scheme 
\begin{align*}
p^{k+1}_\textbf{iN} & = \sum_{i=0}^{k+1} K^i \hat{\textbf{A}}^{-1}_\angle g_\angle 
= K\sum_{i=0}^k K^i \hat{\textbf{A}}^{-1}_\angle g_\angle + \hat{\textbf{A}}^{-1}_\angle g_\angle 
= K p_\textbf{iN}^k + \hat{\textbf{A}}^{-1}_\angle g_\angle\\
& = (I- \hat{\textbf{A}}^{-1}_\angle \hat{\textbf{A}}_\angle) p^k_\textbf{iN} + \hat{\textbf{A}}^{-1}_\angle g_\angle \\
& = p^k_\textbf{iN} - \hat{\textbf{A}}^{-1}_\angle(\hat{\textbf{A}}_\angle p^k_\textbf{iN} - g_\angle).
\end{align*}}
The effect of the limited-angle PAT Fourier forward model in data domain is demonstrated in \cref{image:BallsFB} on the 4-disk phantom with line sensor on top. The disks 
are placed inside the north sector of the square domain so to eliminate any effect of the finite sensor. 
%
As the full data Fourier forward operator \cref{eq:FrequencyEquation} is not well defined due to the blow up of the factor for the frequencies approaching the boundary of the open range (large bow-tie in \cref{image:BowtieTheta60} (b)), we have to limit the factor which corresponds to limiting the sensitivity angle away from $\pi/2$ (albeit only by $\epsilon>0$, very small). Thus the full angle is chosen as $\pi/2 - \epsilon$. The nearly full-angle data ($\theta_\text{max} = \pi/2 - \epsilon, \, \epsilon > 0$, small) shown in (b) is dominated by slanted lines. These slanted lines align with the wavefront mapping of $\pi/2-\epsilon$ and make an angle $\arctan(\sin(\pi/2-\epsilon)) \approx \pi/4$ with the sensor. We suppose that the limited, but still large factor, in the nearly full-data forward operator has the effect of adding these slanted singularities, akin to the singularities added by a sharp data cut-off in limited-view scenarios.
Furthermore, the slanted lines wrap around along the horizontal dimension due to rectangular aspect ratio of the domain (the represented vertical frequencies are higher than the represented horizontal frequencies). 
For the limited-angle data ($\theta_\text{max} = \pi/4$)\footnote{In real data acquisition scenarios with Fabry Perr\`ot sensor, $\theta_{\max}$ is empirically found to be $\approx\pi/4$.} in (c) these artefacts are significantly reduced as the narrower bow-tie 
i) effects a stronger limit on the factor, ii) is effectively a high frequency cut-off resulting in a smoothing of the time-space domain data. Smoothing is a common strategy to reduce artefacts due to singularities added by the sharp cut-off of time-space domain data (e.g.~due to the finite size sensor). This is consistent with microlocal analysis and was also proposed by Frikel and Quinto in \cite{frikelQuinto2015_limitedViewTomoPAT} for limited sensor PAT.
While there is an apparent similarity between (d) the limited-angle inversion ${p_0}^\text{Linear}=\hat{\textbf{A}}^{-1}_\angle g_\angle$ and (e) the limited-angle adjoint $p_0^\text{Adj}=\hat{\textbf{A}}^{*}_\angle g_\angle$, we point out the noticeable difference in contrast (s.a.~(f) for the visualisation of their difference) due to the frequency dependent factor $|\bk|/|\bk_{\perp}|$ present in the inverse but not in the adjoint.

A peculiarity of the half space geometry with planar/line sensor, is that each wavefront is measured exactly once i.e.~we measure the wavefront travelling towards the detector and the wavefront in the opposite direction is accounted for by assuming the mirror symmetry of the data w.r.t.~the detector hyperplane. Thus neither the limited sensitivity angle nor the limited sensor result in any partially measured singularities (all or nothing) and hence no improvement on linear reconstruction can be expected by a Neumann series type approaches.

\dontshow{We observe that the Neumann series reconstruction $\textbf{iN}(g_\angle)$ (f) is almost the same as the direct linear inversion ${p_0}^\text{Linear}$ (d). This is a consequence of the half space geometry with planar / line sensor, where each wavefront is measured exactly once i.e.~we measure the wavefront travelling towards the detector and the wavefront in the opposite direction is accounted for by assuming the mirror symmetry of the data w.r.t.~the detector hyperplane. Thus the limited sensitivity angle (nor actually the limited sensor) does not result in any partially measured singularities (all or nothing) and hence no improvement can be expected by the iterative Neumann series approach.}

\section{Multiscale Representation of Photoacoustic Initial Pressure}
\label{sec:CurveletRepresentationImage}
In this section we discuss multiscale representations used throughout the paper. We start with the standard Curvelet transform 
and introduce its restriction to the wedge of directions, fully wedge restricted Curvelet transform.  Fully wedge restricted Curvelet transform can provide a microlocal properties preserving representation of either the null space of the discrete limited-angle operator (invisible directions) or its complement (visible directions), depending of the choice of the wedge. Finally, we introduce a Coronae decomposition which is effectively the scale (frequency band) decomposition underpining the Curvelet transform used in Section \ref{sec:sec:sec:CoroaneNet} to construct a U-Net with a multiscale structure matching this of the Curvelet frame.

\subsection{Curvelets}
\label{sec:sec:Curvelets}
The Curvelet transform \cite{candes2006fast} is a multiscale pyramid with many directions and positions at each scale. 
\cref{image:Tiling} (a) shows the Curvelet induced tiling of the Fourier domain in 2D, with Curvelet window functions 
supported near a trapezoidal wedge with the orientation $\theta_l$, $\theta_l \in [-\pi, \pi)$ and the scale $j$. The corresponding Curvelet envelope function in spatial domain 
is aligned along a ridge of length $2^{-j/2}$ and width $2^{-j}$.
The wedges/envelopes become finer with increasing scales which lends the Curvelet frame the ability to resolve singularities along curves \cite{candes2005continuous}, \cite{candes2004new}, \cite{candes2006fast}.
\dontshow{
\begin{figure}[htbp]
\centering    
\includegraphics[width=0.65\linewidth]{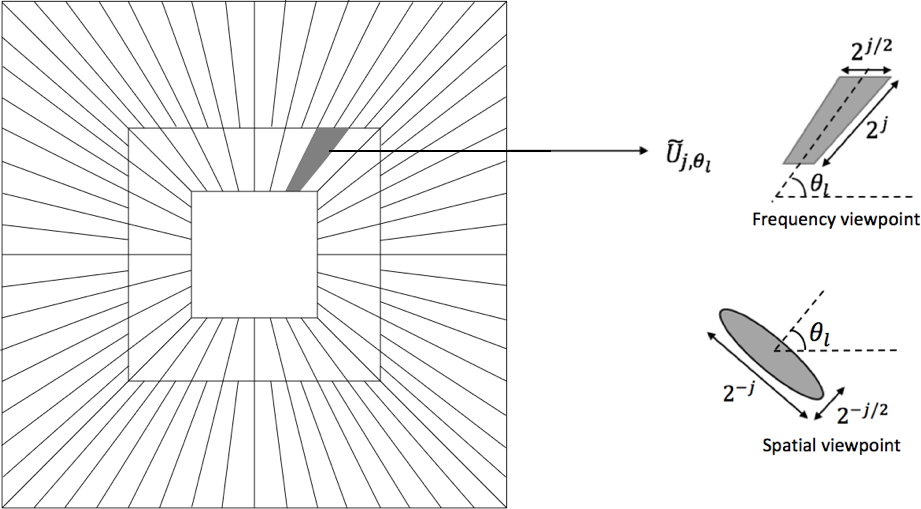}
\caption[2DTilling]{An example of 2D frequency domain Curvelet tiling using 3 scales. The highlighted wedge, magnified view in top right, corresponds to the
frequency window near which the Curvelet $\tilde{U}_{j,\theta_l}$
at the scale $j$ with orientation $\theta_l$ is supported. The orientation of the envelope of the corresponding Curvelet function in the spatial domain is shown in bottom right corner.}
\label{image:2DTiling}
\end{figure}}

We introduce the Curvelet transform following the continuous presentation in \cite{candes2006fast}. For each scale $j$, orientation $\theta_l =  2\pi/L \cdot 2^{-\lfloor j/2 \rfloor} \cdot l , l \in \mathbb Z^{d-1}$ such that $\theta_l \in [-\pi,\pi)^{d-1}$, where $L \cdot 2^{\lfloor j_0/2 \rfloor}$ is the number of angles at the second coarsest scale $j=j_0+1$, $j_0$ even, the Curvelet coefficients of $u: \mathbb{R}^d\rightarrow \mathbb R$, are computed as 
\begin{equation}\label{eq:NonwrapCurveletformula}\tag{\textbf{C}}
C_{j,l}({\bf a}) = \int_{\mathbb R^d} \hat{u}({\bf k})\tilde{U}_{j,\theta_l}({\bf k}) e^{i {\bx_{\ba}^{(j,l)}} \cdot \bk} d{\bf k},
\end{equation}
which is a projection of the Fourier transform $\hat{u}(\mathbf{k})$ on the corresponding frame element function $\tilde{U}_{j,\theta_l}(\bk) \exp(-i\bk \cdot \bx_{\ba}^{(j,l)})$ with $\tilde{U}_{j,\theta_l}({\bf k})$ a trapezoidal frequency window with scale $j$ and orientation $\theta_l$, and a spatial domain centre at $\bx_{\ba}^{(j,l)} = (a_1 \cdot 2^{-j}, a_2 \cdot 2^{-j/2}, \dots, a_d \cdot 2^{-j/2}) \in \mathbb R^d$ corresponding to the sequence of translation parameters $\ba = (a_1,\dots, a_d)\in \mathbb{Z}^d$. Note that the coarse scale Curvelets are isotropic Wavelets corresponding to translations of an isotropic low pass window $\tilde{U}_{0}({\bf k})$. The window functions at all scales and orientations form a smooth partition of unity on the relevant part of $\mathbb{R}^d$. For real function $u$, we use the symmetric version of the Curvelet transform corresponding to the Hermitian symmetry of the wedges $\theta_l$ and $\theta_l+\pi$ in the frequency domain. There are two different ways to evaluate the integral \cref{eq:NonwrapCurveletformula} efficiently. In this paper, we use the implementation via wrapping i.e.~the digital coronisation with shears introduced in \cite{candes2006fast}. This fast digital Curvelet transform is a numerical isometry. We refer to \cite{candes2006fast} for further details and to Curvelab package\footnote{http://www.curvelet.org/software.html} for the 2D and 3D implementations.

\subsection{Fully Wedge Restricted Curvelets}
\label{sec:sec:FWRCurvelets}
Wedge restriction of Curvelet transform was introduced in \cite{pan2021photoacoustic} and we refer to that paper and the accompanying codes\footnote{\url{https://github.com/BolinPan/Wedge_Restricted_Curvelet}} for the details of the general construction in $\mathbb R^d$. Here we restrict the presentation to the symmetric version in 2D as it is sufficient to explain the idea of full wedge restriction and its generalisation to $d > 2$ follows the same principles as the generalisation of the wedge restriction to higher dimensions.
%
%

\begin{figure}[htbp]
\centering
  \subfloat[]{
  \includegraphics[width=0.35\linewidth,height=0.2\linewidth]{Images/2DCurvelet.png}
}
\hspace{1cm}
  \subfloat[]{
  \includegraphics[width=0.2\linewidth,height=0.2\linewidth]{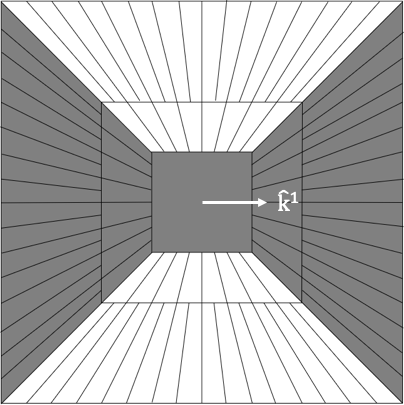}
}
  \subfloat[]{\includegraphics[width=0.2\linewidth,height=0.2\linewidth]{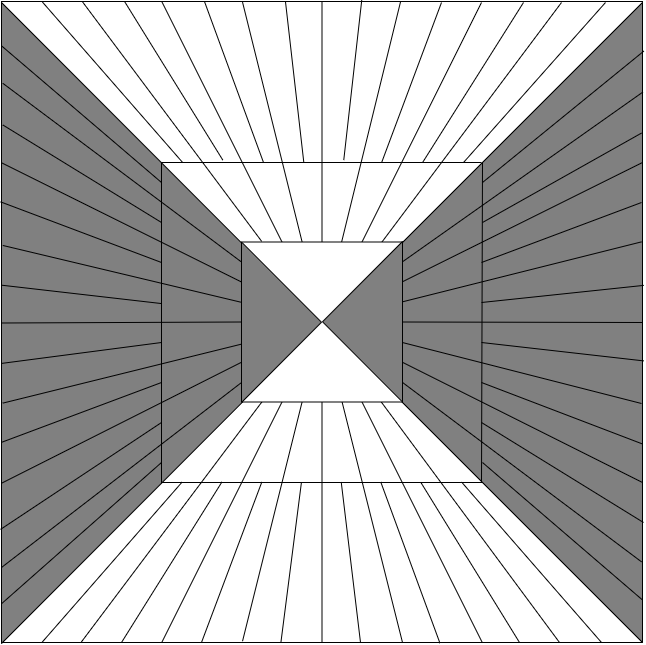}}
  \caption[2Dgrey]{(a) An example of a 2D frequency domain Curvelet tiling using 3 scales. The highlighted wedge, magnified view in top right, corresponds to the frequency window  $\tilde{U}_{j,\theta_l}$ at scale $j$ with orientation $\theta_l$ near which a Curvelet is supported. The orientation of the envelope of the corresponding Curvelet function in the spatial domain is shown in bottom right corner. Induced (b) wedge  restricted (c) fully wedge restricted Curvelet transforms for $\theta_{\max} = \pi/4$. The respective bow-tie shaped ranges are coloured \textcolor{gray}{gray}.} 
  \label{image:Tiling}
\end{figure}

The 2D \emph{symmetric wedge restricted Curvelet transform} restricts the Curvelet orientations to the symmetric double wedge (bow-tie) $\theta_l \in W := [-\theta^w,\theta^w] \cup [-\theta^w+\pi,\theta^w+\pi]$ with 
$\theta^w \in (0,\pi/2)$
as illustrated in \cref{image:Tiling} (b), 
where the gray region corresponds to Curvelets with orientations inside the bow-tie range. 

\dontshow{Formally, we can write the wedge restricted Curvelet transform $\tilde \Psi : \mathbb R^{n} \rightarrow \mathbb R^{\tilde N}$, $\tilde N<N$ as a composition of the standard Curvelet transform $\Psi : \mathbb R^{n} \rightarrow \mathbb R^{N}$ and a projection operator $P_W : \mathbb R^{N} \rightarrow \mathbb R^{\tilde N}$, $\tilde{\Psi} = P_W \Psi$:
\begin{align}
\label{eq:projR}
&P_W: \mathbb{R}^N \rightarrow \mathbb{R}^{\tilde N}\\
\nonumber &P_W(C_{j,l}(\mathbf{a})) =
\begin{cases}
C_{j,l}(\mathbf{a}),  & \theta_{j,l}\in W, \; j > j_0\\
C_{j,l}(\mathbf{a}),  & j = j_0\\
\mathbf{0},  & \text{otherwise},
\end{cases}
\end{align}
where the angles $\theta_{j,l}$ are the discrete wedge orientations (correspond to their centres).
At the coarse scale, $j=j_0$, Curvelets are isotropic Wavelets, they are non-directional and thus are mapped to themselves. \bp{check the correctness of this statement:}We note the fully wedge restricted Curvelet transform is an isometry when defined on the restriction to the range of $P_W$ i.e.~${\tilde \Psi}\, : \left.\mathbb R^n\right|_{\textrm{range}(P_W)} \rightarrow \mathbb R^{\tilde N}$, where $\textrm{range}(P_W)$ corresponds to the set $W \cup \tilde U_{0}$ in the Fourier domain.}

The drawback of the original formulation of the wedge restricted Curvelet transform are the isotropic Wavelet window functions at the coarse scale inherited from the original Curvelet transform which result in non-directional coarse scale coefficients. 
To overcome this shortcoming, we enforce the 
split into in and out of wedge frequencies
also at the coarse scale $j_0$ 
\dontshow{
\todo{Bolin" which one is it: 1) bow-tie to the coarse scale image in Fourier domain (coarse scale as we compute it in Cornae coefficients) or bow tie to the Curvelet coefficients which I think also is an image but the pixels correspond to smooth bump filters and then inverse FFT each separately. As the wedge filters at higher scales are smooth, our restriction for higher scales is smooth and 2) is more compatible. Choose: 
by applying a binary bow-tie filter in the Fourier domain to the coarse scale, 
applying a binary bow-tie filter to the coarse scale isotropic Curvelet coefficients,} \bp{I applied the bow-tie to the Curvelet coefficients in its Fourier domain i.e. smooth bump filters then bow-tie filter at coarse scale, then inverse FFT to get the coarse scale coefficient (which is an image) as shown in Figure 5 (f).}
}
applying a binary bow-tie filter to the coarse scale isotropic Curvelet coefficients\footnote{The isotropic Curvelet coefficients are projections on the translations of the smooth isotropic low pass window $\tilde U_0(\bk)$. Thus our construction results in a smoothed bow-tie restriction, however, all the frequency windows and higher scales are also smooth, thus smooth restriction is applied consistently to all scales.},
and henceforth we refer to this construction as a \emph{fully wedge restricted Curvelet transform}\footnote{The code is available from \url{https://github.com/BolinPan/CoronaeNet}}. 

Formally we can write the fully wedge restricted Curvelet transform  $\Breve{\Psi} : \mathbb R^{n} \rightarrow \mathbb R^{\Breve N}$, $\Breve N < N$ as a composition of the standard Curvelet transform $\Psi : \mathbb R^{n} \rightarrow \mathbb R^{N}$ and a projection operator $P_W : \mathbb R^{N} \rightarrow \mathbb R^{\Breve N}$, $\Breve{\Psi} = P_W \Psi$
\begin{align}
\label{eq:projCR}
&P_W: \mathbb{R}^N \rightarrow \mathbb{R}^{\Breve N}\\
\nonumber &P_W(C_{j,l}(\mathbf{a})) =
\begin{cases}
C_{j,l}(\mathbf{a}),  & \theta_{j,l}\in W, \, j \geq j_0\\
\mathbf{0},  & \text{otherwise},
\end{cases}
\end{align}
where the angles $\theta_{j,l}$ are the discrete wedge orientations (essentially corresponding to the wedge centres) and we extend this notation to the coarse scale $j_0$ as $\theta_{j_0, l} = \textrm{atan2}({\bf x_a}^{(j_0,l)})$. 
We note the fully wedge restricted Curvelet transform is an isometry when defined on the restriction to the range of 
$\Breve\Psi^\dagger$ i.e.~${\Breve\Psi}\,: \left.\mathbb R^n\right|_{\textrm{range}(\Breve\Psi^\dagger)} \rightarrow \mathbb R^{\Breve N}$, where $\textrm{range}(\Breve\Psi^\dagger)$ corresponds to the set $W$ in the Fourier domain. 
\cref{image:Tiling} (b,c) illustrate in Fourier domain the difference between the ranges of the wedge restricted and fully wedge restricted 2D Curvelet transforms.

\subsection{Fully Wedge Restricted Curvelet Representation of Initial Pressure}
\label{sec:sec:FWRC}
Curvelets provide an almost optimally sparse representation if the function $u: \, \mathbb R^d \rightarrow \mathbb R$ is smooth away from (piecewise) $\mathcal{C}^2$ singularities \cite{candes2004new, starck2002curvelet, donoho2001sparse}. More precisely, the error of the best $s$-term approximation, $u_s$, (corresponding to taking the $s$ largest in magnitude coefficients) in Curvelet frame decays as \cite{candes2004new}
\begin{equation}
||u-u_s||^2_2 \leq \mathcal{O} ( (\log s)^3\cdot s^{-2} ). 
\end{equation}
Furthermore, Curvelets have been shown to sparsely propagate through the wave equation \cite{candes2005curvelet} essentially being translated along the trajectories of the Hamiltonian flow with initial conditions corresponding to the spatial domain centre and the direction of the Curvelet. For constant speed of sound, this implies that the angle $\theta$ at which the Curvelet impinges on the detector is the same as the Curvelet orientation in the decomposition of the initial pressure $p_0$. A direct consequence of the restriction of the angle $|\theta| \leq \theta_{\max}$ is the restriction to the \emph{visible} initial pressure $p_0$, which corresponds to symmetric fully wedge restricted Curvelet transform with directions $\theta_l \in W^\theta_{\max}, \, W^\theta_{\max} := [-\theta_{\text{max}},\theta_{\text{max}}] \cup [-\theta_{\text{max}}+\pi,\theta_{\text{max}}+\pi]$ with 
$\theta_{\text{max}} \in (0, \pi/2)$.
On the other hand, the restriction of the impingement angle, $|\theta| \leq \theta_{\max}$, translates via \cref{eq:WaveFrontMapping} into the restriction of the angle $\beta$, $|\beta| \leq \arctan(\sin\theta_\text{max})$, implying narrowing of the bow-tie shaped range of the PAT forward operator to $W^\beta_{\max} := [-\beta_{\max},\beta_{\max}] \cup [-\beta_{\max} +\pi,\beta_{\max}+\pi]$ with 
$\beta_{\max} = \arctan(\sin\theta_{\max}) \in (0, \pi/4)$.

We now reinterpret \cref{eq:NonwrapCurveletformula} to obtain the symmetric fully wedge restricted Curvelet transform for representation of the \emph{visible} initial pressure 
\begin{align}\label{eq:Cp0}\tag{\textbf{FWRC}}
\Breve{C}_{j,l}({\bf a}_{\perp}, {\bf a}_{\mathcal S}) =\iint_{\mathbb R^d} \hat{u}\left(\bk_{\perp}, \bk_{\mathcal S}\right) \tilde{U}_{j,\theta_{j,l}}\left(\bk_{\perp}, \bk_{\mathcal S} \right) 
e^{i (\bx_{\bf a} \cdot (\bk_{\perp}, \bk_{\mathcal S}) )} 
d\bk_{\perp} d{\bk_{\mathcal S}}, \quad \theta_{j,l} \in W_{\max},
\end{align}
with the following notation:
\begin{enumerate}[leftmargin=6em]
    \space \item[$(\bk_{\perp}, \mathbf{k_{\mathcal S}})$] ambient/image domain frequency $\bk = (\bk_{\perp}, \mathbf{k_{\mathcal S}})$;
    \item[$\theta_{j,l}$] a discrete direction of a trapezoidal frequency window $\tilde{U}_{j,\theta_{j,l}}(\bk_{\perp}, \bk_{\mathcal S})$. The tiling is computed using standard Curvelet transform on a cuboid domain, followed by the projection $P_W$ on the bow-tie shaped set $W_\text{max}$;
    We remark, that efficient implementation would bypass the computation of the wedges outside $W_\text{max}$, but it would require modifications to the Curvelet Toolbox functions.
    \item[$W_\text{max}$] the restriction of $\theta_{j,l}$ to those in the bow-tie $W_\text{\text{max}} := [-\theta_{\text{max}},\theta_{\text{max}}] \cup [-\theta_{\text{max}}+\pi,\theta_{\text{max}}+\pi]$ with $\theta_{\text{max}} \in (0, \pi/2)$ 
    effects the projection on the \emph{visible} range of initial pressure in the Fourier domain;
    \item[$({\bf a}_{\perp}, {\bf a}_{\mathcal S})$] the grid of spatial translations: ${\bf a}_{\mathcal S}$ parallel to the detector $\mathcal S$,  ${\bf a}_{\perp}$ perpendicular to $\mathcal S$ (note that Curvelet transform via wrapping uses one grid per each quadrant at each scale, a.k.a.~$\bx_{\bf a}^{j,\mathbb Q} = (2^{-j}{\bf a}_{\perp}^{\mathbb Q}, 2^{-j/2}{\bf a}_{\mathcal S}^{\mathbb Q}), \, \mathbb Q  = \{\mathbb N, \mathbb W, \mathbb S, \mathbb E\}$). 
\end{enumerate}

We note that in practice, the function $p_0 : \mathbb R^d \rightarrow \mathbb R$ is discretised on an $n$ point grid in
$\mathbb R^d$, yielding a vector $p_0 \in \mathbb R^n$, and we apply the symmetric discrete Curvelet
transform $\Psi\, : \,  \mathbb R^n \rightarrow \mathbb R^N$. The discrete counterparts of the wedge restricted $\tilde\Psi$ and fully wedge restricted $\Breve\Psi$ transforms follow analogously.

\dontshow{
This results in the following for fully wedge restricted Curvelet coefficients
\begin{align}\label{eq:Cp0}\tag{\textbf{FWRC}}
\Breve{C}_{j,l}({\bf a}_{\perp}, {\bf a}_{\mathcal S}) =\iint_{\mathbb R^d} \hat{u}\left(\bk_{\perp}, \bk_{\mathcal S}\right) \tilde{U}_{j,\theta_{j,l}}\left(\bk_{\perp}, \bk_{\mathcal S} \right) 
e^{i (\bx_{\bf a} \cdot (\bk_{\perp}, \bk_{\mathcal S}) )} 
d\bk_{\perp} d{\bk_{\mathcal S}}
\end{align}
with $\bx_{\bf a} = (2^{-j}{\bf a}_{\perp}, 2^{-j/2}{\bf a}_{\mathcal S})$, $\theta_{j,l} \in W_\text{max}$.\par
}

\begin{figure}[htbp!]
\centering
  \subfloat[$|\Psi p_0|$]{
  \includegraphics[width=0.18\linewidth,height=0.18\linewidth]{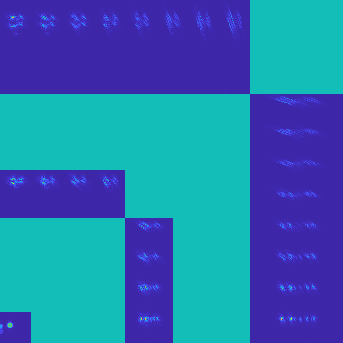}}
  \subfloat[$|\tilde\Psi p_0|$]{
  \includegraphics[width=0.18\linewidth,height=0.18\linewidth]{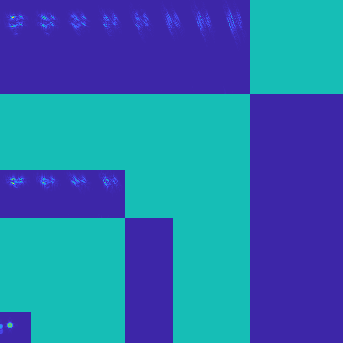}}
  \subfloat[$|\Breve{\Psi} p_0|$]{
  \includegraphics[width=0.18\linewidth,height=0.18\linewidth]{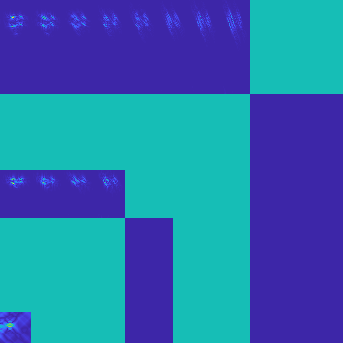}}
  \\
  \subfloat[$|\Psi p_0|_\text{C}$]{
  \includegraphics[width=0.18\linewidth,height=0.18\linewidth]{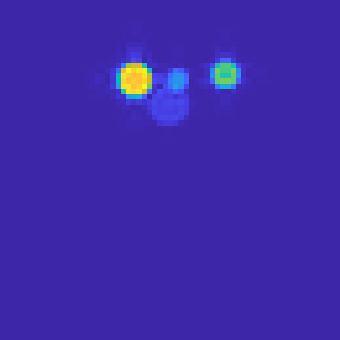}}
  \subfloat[$|\tilde\Psi p_0|_\text{C}$]{
  \includegraphics[width=0.18\linewidth,height=0.18\linewidth]{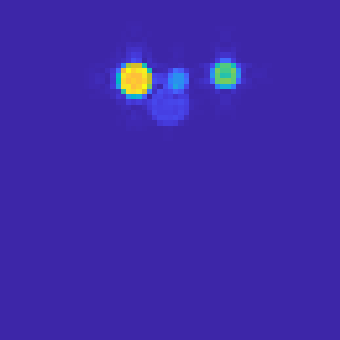}}
  \subfloat[$|\Breve{\Psi} p_0|_\text{C}$]{
  \includegraphics[width=0.18\linewidth,height=0.18\linewidth]{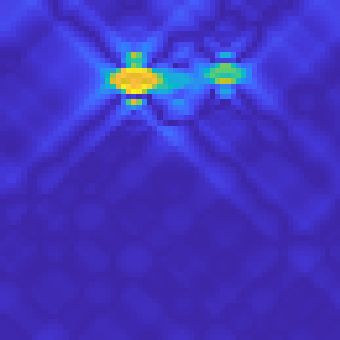}}
  
  \caption[BallsCoeffs]{4-disk Phantom - Curvelet coefficients of 4-disk phantom ($\theta_{\text{max}} = \pi/4$ for wedge restriction): (a) standard, (b) wedge restricted, (c) fully wedge restricted Curvelet transform coefficient magnitudes (showing top-right quarter of scale-normalised coefficient magnitudes); (d-f) magnitudes of the} corresponding coarse scale coefficients.
  \label{image:BallsCoeffs}
\end{figure}

We compare the standard Curvelet transform $\Psi$, wedge restricted Curvelet transform $\tilde \Psi$ with fully wedge restricted Curvelet transform $\Breve{\Psi}$ for $\theta_\text{max} = \pi/4$ on the 4-disk phantom (shown in \cref{image:BallsFB} (a)). We use 3 scales and 32 angles (at the 2nd coarsest level) for the standard Curvelet transform, yielding wedge restricted transforms with $16$ angles (at the 2nd coarsest level). Both the wedge restricted Curvelet transform and the fully wedge restricted Curvelet transforms exclude the wedges out of the bow-tie range $W_\text{max}$ at the higher $j > j_0$ scales; see \cref{image:BallsCoeffs} (b,c). The standard Curvelet transform and the wedge restricted Curvelet transform have the same coarse scale coefficients while the fully wedge restricted Curvelet transform enforces the split of the coarse scale into visible $\theta_{j,l} \in W_{\max}$ and invisible $\theta_{j,l} \not \in W_{\max}$ to obtain the visible initial pressure only; see \cref{image:BallsCoeffs} (d-f). 

\begin{figure}[htbp!]
\centering
  \subfloat[$\Psi^{\dagger} \Psi p_0$]{
  \includegraphics[width=0.21\linewidth,height=0.18\linewidth]{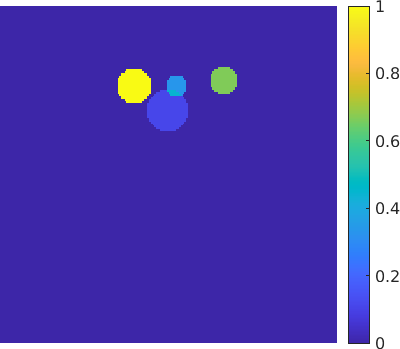}}
  \subfloat[$\tilde\Psi^\dagger \tilde\Psi p_0$]{
  \includegraphics[width=0.21\linewidth,height=0.18\linewidth]{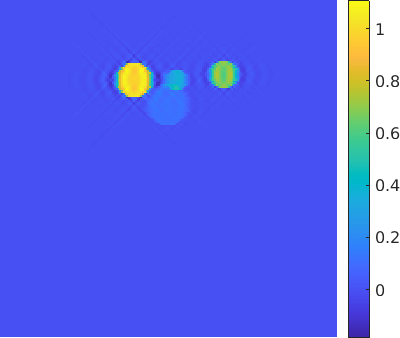}}
  \subfloat[$\Breve{\Psi}^\dagger \Breve{\Psi} p_0$]{
  \includegraphics[width=0.21\linewidth,height=0.18\linewidth]{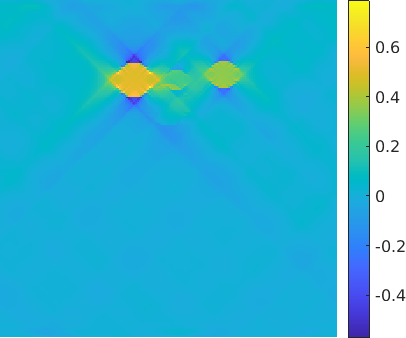}}
  \\
  \subfloat[$\Psi^{\dagger} \Psi p_0 - p_0$]{
  \includegraphics[width=0.22\linewidth,height=0.19\linewidth]{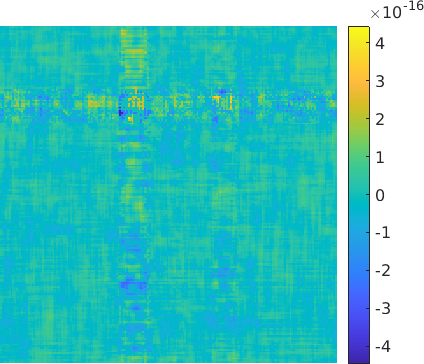}}
  \subfloat[$\tilde\Psi^\dagger \tilde\Psi p_0 - p_0$]{
  \includegraphics[width=0.21\linewidth,height=0.18\linewidth]{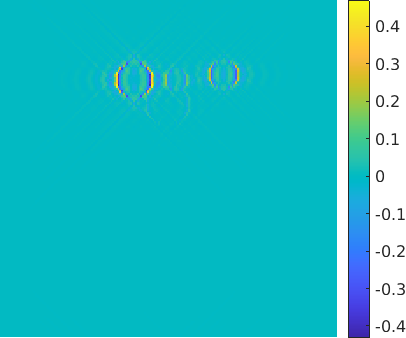}}
  \subfloat[$\Breve{\Psi}^\dagger \Breve{\Psi} p_0 - p_0$]{
  \includegraphics[width=0.21\linewidth,height=0.18\linewidth]{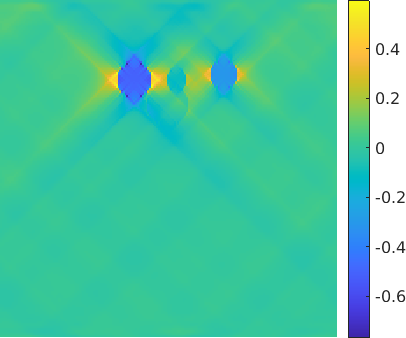}}
  \caption[BallsRec]{4-disk Phantom - Visualisation of the image domain projections effected by the various Curvelet transforms ($\theta_{\text{max}} = \pi/4$ for wedge restriction): 
  (a) standard $\Psi$, (b) wedge restricted $\tilde\Psi$, (c) fully wedge restricted $\Breve{\Psi}$; (d-f) projections on the corresponding orthogonal complements in $\mathbb R^d$.}
  \label{image:BallsRec}
\end{figure}

The image domain effect of projection induced by all three Curvelet transforms are shown in \cref{image:BallsRec}. \cref{image:BallsRec} (c,f) illustrates that the projection 
corresponding to the
fully wedge restricted Curvelet transform corresponds to the visible initial pressure.

\subsection{Coronae Decomposition and Reconstruction}
\label{sec:sec:sec:CoronaeDR}
The Coronae decomposition is effectively the scale (frequency band) decomposition underpinning the Curvelet transform.
Coronae decomposition is computed in the Fourier domain but the Coronae coefficients at each scale are images obtained with inverse scale restricted Fourier transform. We make use of the same smooth partition of unity filters as used in Curvelet Toolbox. 

We henceforth denote the filters at the scale $j$ with $\mathbb{L}_j: \mathbb C^{n_j \times m_j} \rightarrow \mathbb C^{n_{j-1} \times m_{j-1}}$ the low-pass filter followed by the restriction to the range $\mathbb C^{n_{j-1} \times m_{j-1}}$, with $\mathbb{L}^\wedge_j: \mathbb C^{n_j \times m_j} \rightarrow \mathbb C^{n_{j} \times m_{j}}$ the same filter but upsampled to $\mathbb C^{n_j \times m_j}$ via zero padding (equivalent to $\mathbb{L}_j$ without the range restriction)  and with $\mathbb{H}_j: \mathbb C^{n_j \times m_j} \rightarrow \mathbb C^{n_{j} \times m_{j}}$, the high-pass filter which completes the partition of unity $\mathbb{H}_j = \sqrt{1 - (\mathbb{L}^{\wedge}_j)^2}$. Here $n_j \times m_j$ is the size of the image $p_j \in \mathbb R^{n_j \times m_j}$ at the scale $j$ 
\begin{equation}
n_j \times m_j = \left( \frac{n_J}{3\cdot 2^{J-j-1}} + 2\cdot  \lfloor \frac{n_J}{3\cdot 2^{J-j}} \rfloor +1 \right) \times \left( \frac{m_J}{3\cdot 2^{J-j-1}} + 2\cdot \lfloor \frac{m_J}{3\cdot 2^{J-j}} \rfloor+1 \right), 
\end{equation}
for $j_0\leq j < j_0 + J$ and $n_J \times m_J$ is the size of the original image, $p_{j_0+J}$.

\begin{figure}[htbp!]
\centering
\subfloat[$p_0$]{
\includegraphics[width=0.21\linewidth,height=0.18\linewidth]{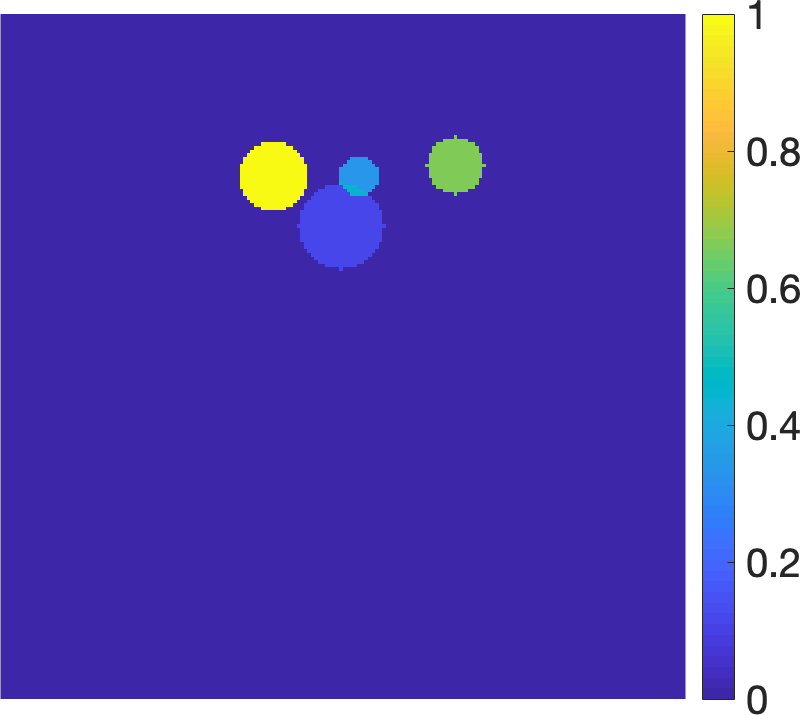}}
\subfloat[$\mathbb{H}_{\rm F} (p_0)$]{
\includegraphics[width=0.21\linewidth,height=0.18\linewidth]{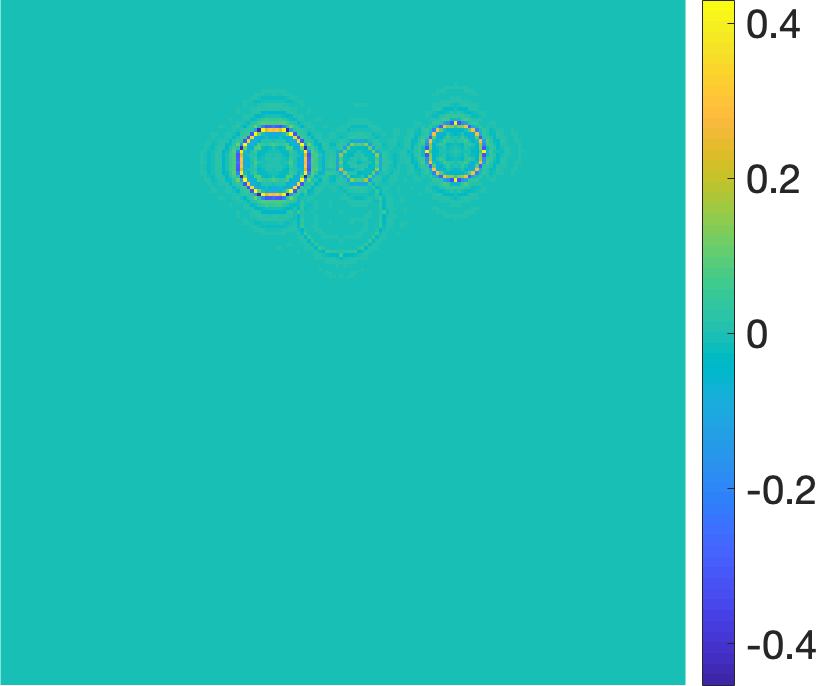}}
\subfloat[$\mathbb{L}_{\rm F} (p_0)$]{
\includegraphics[width=0.17\linewidth,height=0.15\linewidth]{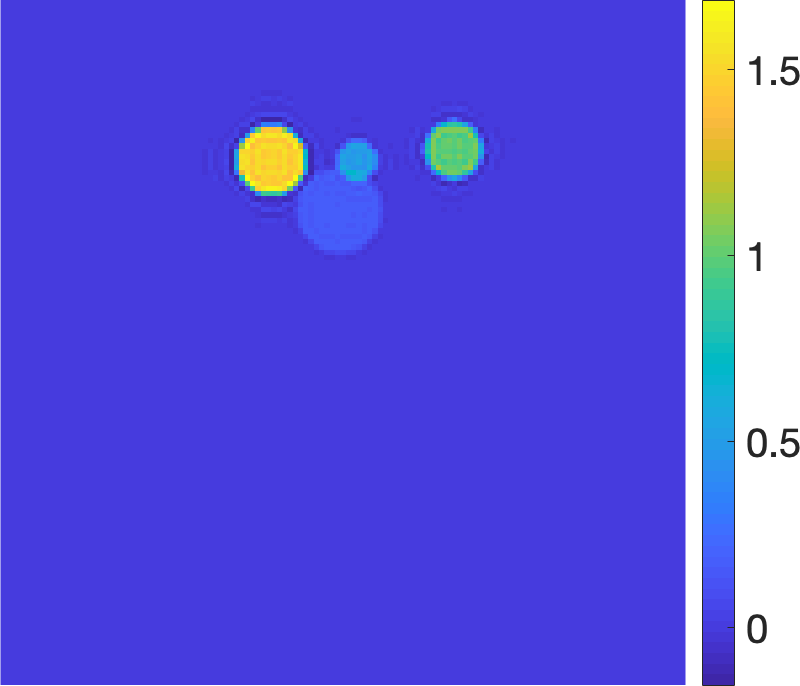}}
\caption{1-level Coronae decomposition of 4-disk phantom: (a) $p_0$, (b) high-pass component $\mathbb{H}_{\rm F}(p_0)$, and (c) low-pass component $\mathbb{L}_{\rm F}(p_0)$.}
\label{image:CoronaeDRImage}
\end{figure}

The Coronae decomposition is effected by application of the filter pair ($\mathbb{L}_j, \mathbb{H}_j$): $\mathbb{H}_j(p_j)$ returns the high-pass component with the same size as $p_j$ and $\mathbb{L}_j(p_j)$ the low-pass component with the size of $p_{j-1}$; see \cref{image:CoronaeDRImage} for visualisation of the effect of 1-level of Coronae decomposition of the 4-disk phantom. 

The signal can be recovered by the Coronae reconstruction. The Coronae reconstruction upsamples the low-pass component via zero-padding and sums up the padded low-pass component with the corresponding high-pass component in the Fourier domain to recover the original signal (denoted with $\oplus$). 

\begin{figure}[htbp!]  
\centering    
\subfloat{
\includegraphics[width=0.4\linewidth,height=0.25\linewidth]{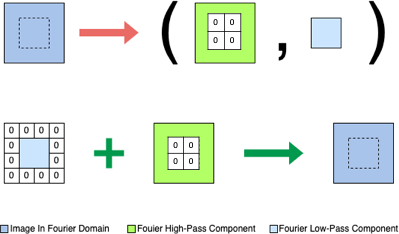}}
\hspace{1cm}
\subfloat{
\includegraphics[width=0.45\linewidth,height=0.25\linewidth,trim={0 0 0 0.001\linewidth},clip]{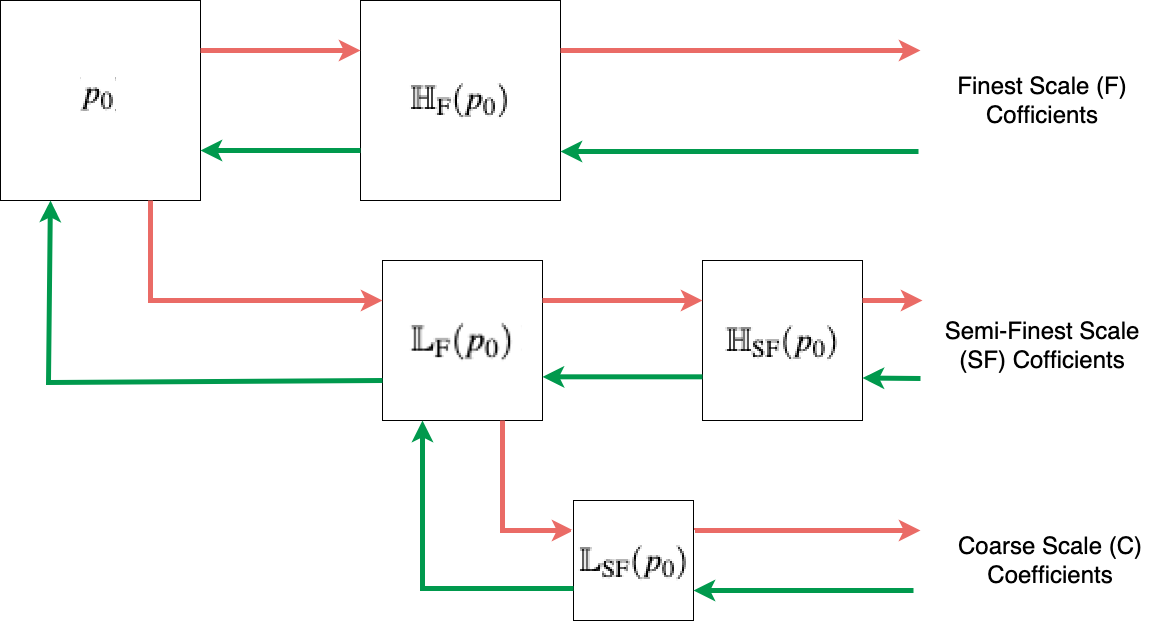}}
\caption{Left: 1-level Coronae decomposition in 2D Fourier domain. Coronae decomposition (top), and reconstruction using zero padding (bottom). Note, the interior of the green square in Fourier domain is populated with 0s}. Right: a filter bank with 3 scales effecting 2-level Coronae decomposition (\textcolor{red}{$\rightarrow$}) and reconstruction (\textcolor{teal}{$\leftarrow$}).
\label{image:CoronaeDRFourier}
\end{figure}

\cref{image:CoronaeDRFourier} (left) show the schematic of the Fourier domain computations involved in Coronae decomposition (top row) and reconstruction (bottom row). For visualisation purposes the overlapping smooth partition of unity filters\footnote{Such filters require appropriate wrapping of the low pass component $\mathbb{L}_j(p_j)$ before Fourier inversion at scale $j-1$.} were replaced with discontinuous non-overlapping partition of unity by indicator functions of the low- and high-band.

Based on the 1-level Coronae decomposition effected by the pair $(\mathbb{L}_j,\mathbb{H}_j)$ we can form a filter bank recursion for multi-level Coronae decomposition
\begin{equation}
    ((\mathbb{L}_{j-1}(\mathbb{L}_j), \mathbb{H}_{j-1}(\mathbb{L}_j)), \mathbb{H}_j), \quad{} j = j_0+J, ..., j_0+1,
\end{equation}
and reconstruction 
\begin{equation}
    (\mathbb{L}_{j} \leftarrow \mathbb{L}_{j-1} \oplus \mathbb{H}_{j-1}) \oplus \mathbb{H}_j,  \quad{} j = j_0+1, ..., j_0+ J,
\end{equation}
\cref{image:CoronaeDRFourier} (right) displays a 3 scale filter bank where the \textcolor{red}{red} and \textcolor{teal}{green} arrows indicate the Coronae \textcolor{red}{decompositions} and \textcolor{teal}{reconstructions}, respectively.
\dontshow{
\begin{figure}[htbp!]  
\centering    
\includegraphics[width=0.7\linewidth]{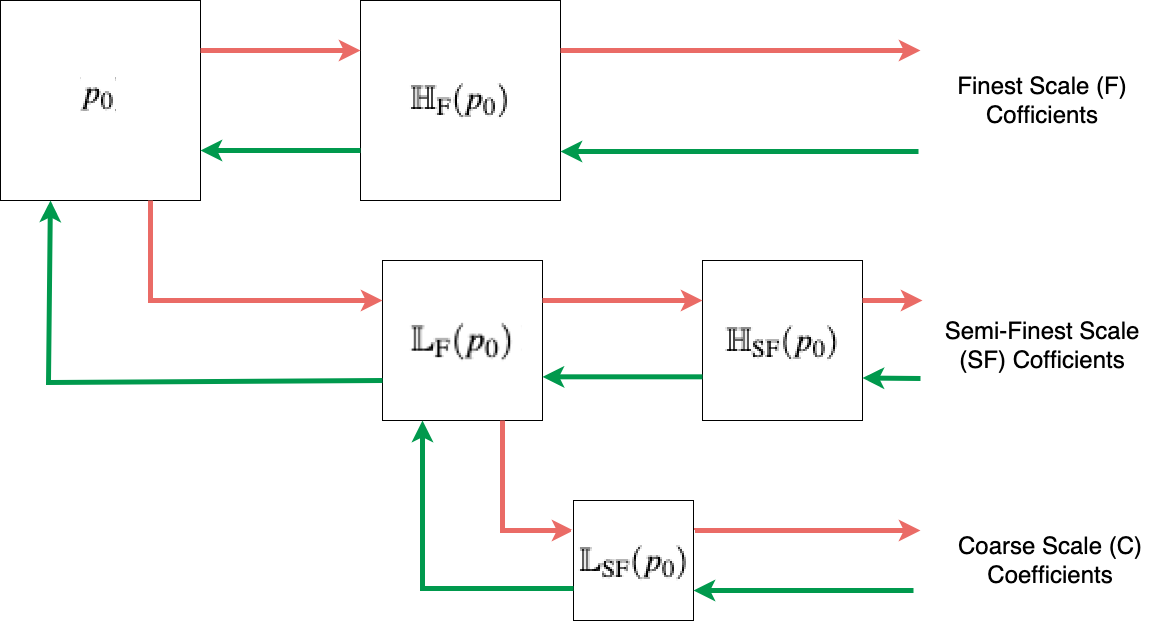}
\caption{A filter bank with 3 scales effecting 2-levels of Coronae decompositions (red arrows) and reconstructions (green arrows).}
\label{image:FilterBank}
\end{figure}}

\subsection{Coronae Decomposition vs Curvelet Decomposition}\label{sec:CoronaeVsCurvelets}
The relation of the Coronae and Curvelet decompositions can be formally stated in the Fourier domain as

\begin{align}\label{eq:CornaeCoeff}\tag{\textbf{QC}}
\mathbb{L}_j\hat Q_j(\bk_j) = \sum_{l}  \sum_{\ba_{j,l}} \tilde U_{j,\theta_l}(\bk_j) C_{j,l}(\ba_{j,l}) e^{-i\bx_\ba^{(j,l)} \cdot \bk_j},
\end{align}
where $\bk_j$ is the frequency domain vector restricted to the scale $j$, $j\geq j_0$\footnote{Note that the overlapping smooth partition of unity filters yield larger $\bk_j$ than the 0-1 non-overlapping filters.} and $\bx_\ba^{(j,l)}$ are the Curvelet centers at scale $j$ and angle $\theta_{j,l}$. In particular, when using implementation via wrapping we have one grid per quadrant $\bx_\ba^{(j,l)} = \bx_\ba^{(j,\mathbb Q(l))}$. 

Applying Coronae transform to the visible / invisible parts of the image (with the maximal sensitivity angle $\theta_{\max}$) corresponds to restricting the sum in \cref{eq:CornaeCoeff} to the visible wedge $l:\, \theta_{j,l} \in W_{\max}$ or to its complement $l:\, \theta_{j,l} \not\in W_{\max}$ for invisible coefficients. This is equivalent to the fully wedge restricted Curvelet transform with the same wedge $W_{\max}$ for visible coefficients and again its complement for invisible coefficients.

\begin{figure}[htbp!]
\centering
\subfloat[$|\Breve{C}_{\text{F}}^\text{vis}|$]{
\includegraphics[width=0.21\linewidth,height=0.18\linewidth]{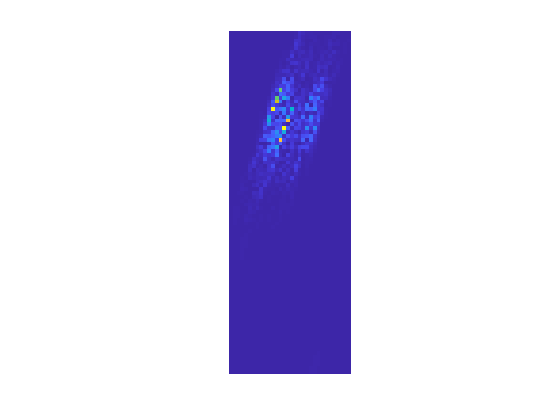}}
  \subfloat[$|\Breve{C}_{\text{SF}}^\text{vis}|$]{
  \includegraphics[width=0.20\linewidth,height=0.16\linewidth]{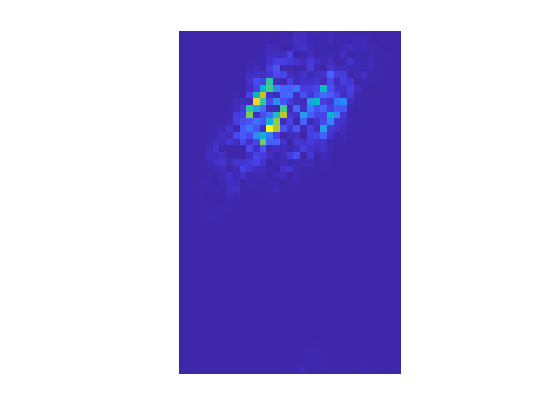}}
  \subfloat[$|\Breve{C}_{\text{C}}^\text{vis}|$]{
  \includegraphics[width=0.18\linewidth,height=0.14\linewidth]{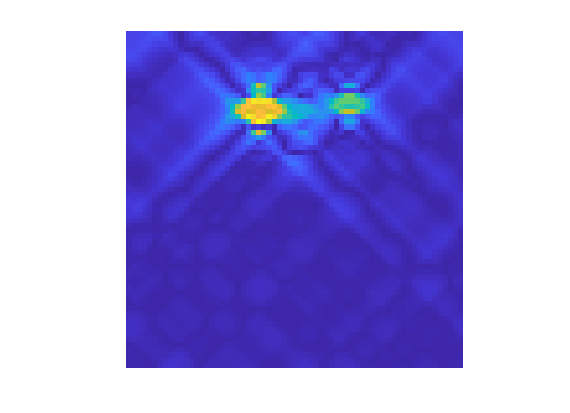}}
\\
\subfloat[$\Breve{\Psi}^\dagger(\Breve{C}_{\text{F}}^\text{vis})$]{
\includegraphics[width=0.21\linewidth,height=0.18\linewidth]{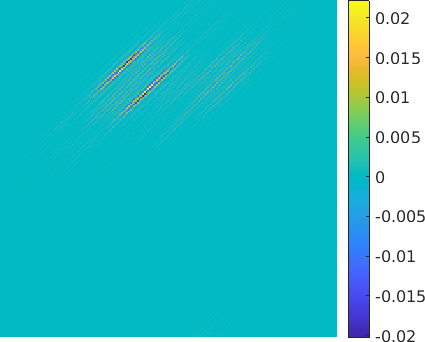}}
  \subfloat[$\Breve{\Psi}^\dagger(\Breve{C}_{\text{SF}}^\text{vis})$]{
  \includegraphics[width=0.20\linewidth,height=0.16\linewidth]{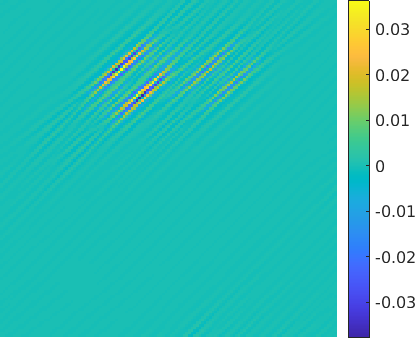}}
  \subfloat[$\Breve{\Psi}^\dagger(\Breve{C}_{\text{C}}^\text{vis})$]{
  \includegraphics[width=0.18\linewidth,height=0.14\linewidth]{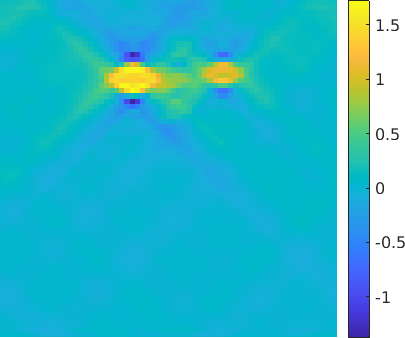}}
\\
  \subfloat[$Q_{\text{F}}^\text{vis}$]{
\includegraphics[width=0.21\linewidth,height=0.18\linewidth]{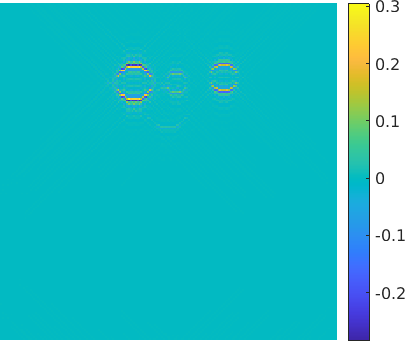}}
  \subfloat[$Q_{\text{SF}}^\text{vis}$]{
  \includegraphics[width=0.20\linewidth,height=0.16\linewidth]{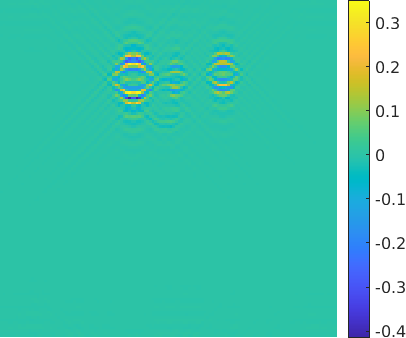}}
  \subfloat[$Q_{\text{C}}^\text{vis}$]{
  \includegraphics[width=0.18\linewidth,height=0.14\linewidth]{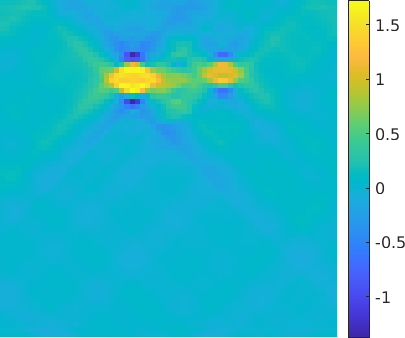}}
  \caption{4-disk Phantom - Representations of the visible ($\theta_{\text{max}} = \pi/4$) with 3 scales $\{\text{C, SF, F}\}$: (a-b) amplitudes of fully wedge restricted Curvelet coefficients in one visible trapezoidal window at scales F, SF (corresponding blue tiles in the Curvelet coefficient display) and (c) amplitudes of all visible coarse scale C coefficients; (d-f) image domain representation of the coefficients in (a-c); (g-i) visible Coronae coefficients at all scales.}
   \label{image:CoronaeCurvelet}
\end{figure}

The computation of Coronae coefficients of the visible/invisible bypasses the computation of the Curvelet transform. After Fourier domain application of the bow-tie shaped filter to extract the visible $\textrm{atan2}(\bk) \in W_{\max}$ and its complement the invisible $\textrm{atan2}(\bk) \not\in W_{\max}$ parts of the image, we use the filters bank defined in \Cref{sec:sec:sec:CoronaeDR} to obtain the Coronae decomposition $Q$ of each visible/invisible component directly. \cref{image:CoronaeCurvelet} (g-i) shows 2-level Coronae decomposition coefficients of the visible part of the 4-disk phantom.



\section{Reconstructing the Visible and Learning the Invisible}
\label{sec:ReconVisLearnInv}
The fully wedge restricted Curvelet transform introduced in \Cref{sec:sec:FWRCurvelets,sec:sec:FWRC}, allows us to decompose the initial pressure $p_0 \in \mathbb R^n$ into Curvelet coefficients that belong to either the visible $f^\text{vis} \in \mathbb R^{\Breve N}$ or the invisible $f^\text{inv} \in \mathbb R^{\Breve N}$ initial pressure (in the sense that the other coefficients are set to 0)
\begin{equation*}\label{eq:VID}\tag{\textbf{VID}}
\begin{aligned}
    f^\text{vis} & = \Breve{\Psi} p_0, \\
    f^\text{inv} & = \Breve{\Psi}^\perp p_0.\\
\end{aligned}
\end{equation*}

Given the limited-angle data $g_\angle \in \mathbb R^m$ contaminated with additive white noise, the \emph{visible} initial pressure $p_0$ can be reconstructed by reconstructing the \emph{visible} coefficients of $p_0$ in the fully wedge restricted Curvelet frame.
The latter is a standard compressed sensing $\ell_2-\ell_1$ minimisation problem with the $\ell_2$ data fidelity term involving the aforementioned limited-angle PAT Fourier forward operator. As the invisible coefficients are 
in the null space of the discrete forward operator i.e.~ $f^\text{inv} \in \textbf{ker}(\hat{\textbf{A}}_\angle)$ \cite{frikel2013sparse}, $f^\text{inv}$ can not be reconstructed from $g_\angle$. Instead, given a representative training set, we can fill in the invisible coefficients using a trained CNN. We refer to such two step method as \emph{reconstructing the visible and learning the invisible}.

\subsection{Reconstructing the Visible}
\label{sec:sec:ReconVis}

In this step, we recover the visible part of initial pressure $p_0 \in \mathbb R^n$ from noisy measurement $g_\angle \in \mathbb{R}^{m}$ in a fully wedge restricted Curvelet frame. To this end we solve 
\begin{equation}\label{VR}\tag{\bf VR}
\tilde{f}^\text{vis} \in \argminA_f \frac{1}{2}||\hat{\textbf{A}}_\angle \Breve{\Psi}^\dagger f-\hat g_\angle||_2^2 + \tau || \Lambda f ||_1,
\end{equation}
where $\hat{\textbf{A}}_\angle : \mathcal D_\textbf{A}^\angle \to \mathcal R_\textbf{A}^\angle$ is the discrete limited-angle PAT Fourier forward operator, $\Breve{\Psi}^{\dagger} : \mathbb{R}^{\Breve N} \rightarrow  \mathbb{R}^{n} $ is the left inverse (and an adjoint)
of the fully wedge restricted Curvelet transform $\Breve{\Psi}$\footnote{As already mentioned in \Cref{sec:PATFourierOperators} 
all the computations 
can be executed directly in the Fourier domain.
For simplicity we use the same notation for transforms ($\Psi$, $\tilde{\Psi}$, $\Breve{\Psi}$), which act on the Fourier transform $\hat p_0$ as for those that act on $p_0$ itself.}, $f$ are the coefficients of $p_0$ in $\Breve{\Psi}$, $\tilde f^\text{vis}$ are the reconstructed visible Curvelet coefficients, and the regularisation parameter $\tau$ is made scale dependent by multiplication with $\Lambda =\diag({2^{{\bf j}-2})}$, where 
${\bf j}\geq j_0 {\bf 1}$ is a vector containing the scales of the frame elements $\Breve{\Psi}$. We propose to solve \cref{VR} using \textbf{FISTA} \cite{FISTA} with a pre-computed Lipschitz constant $L = ||\Breve{\Psi}\hat{\textbf{A}}^{\dagger}_\angle \hat{\textbf{A}}_\angle \Breve{\Psi}^{\dagger}||_2$. 
\dontshow{
A version of \textbf{FISTA} adapted to \cref{VR} is summarized in Algorithm \textbf{VR-FISTA}.
\begin{VR-FISTA} 
\renewcommand{\thealgorithm}{}
\small 
\caption{Fast Iterative Shrinkage Thresholding Algorithm \cite{FISTA}}
\begin{algorithmic}[1]
\STATE \textbf{Initialisation:} $y^1 = \mathbf{0}$, $f^1 = \mathbf{1}$, $\Lambda = \text{diag}(2^{j-2})$, $j\geq j_0$, $\alpha^1 = 1$, $\mu=\frac{1}{L}$, $\tau>0$, $\eta>0$ and $K_{\text{max}}$ \par
\STATE $k := 1$\par
\STATE \textbf{Repeat}\par
\STATE $\tilde{z}^k = y^k - \mu \Breve{\Psi} \hat{\textbf{A}}_\angle^{\dagger} (\hat{\textbf{A}}_\angle \Breve{\Psi}^{\dagger}y^k - \hat g_\angle)$\par
\STATE $f^{k+1} = \argminA_f \mu\tau||\Lambda f||_1+\frac{1}{2}||f-\tilde{z}^k||_2^2$\par
\STATE $\alpha^{k+1} = (1+\sqrt{1+4(\alpha^k)^2})/2$\par
\STATE $y^{k+1} = f^{k+1} + \frac{\alpha^k-1}{\alpha^{k+1}}(f^{k+1} - f^{k})$\par
\STATE $k = k+1$ \par
\STATE \textbf{Until} $||f^{k}-f^{k-1}||_2 /||f^{k-1}||_2<\eta \text{ or } k>K_{\text{max}}$
\end{algorithmic}
\label{algo:VRFISTA}
\end{VR-FISTA}
}

\subsection{Learning the Invisible}
\label{sec:sec:learnInv}
\begin{figure}[htbp!]
\centering    
\includegraphics[width=0.8\linewidth,height=0.45\linewidth]{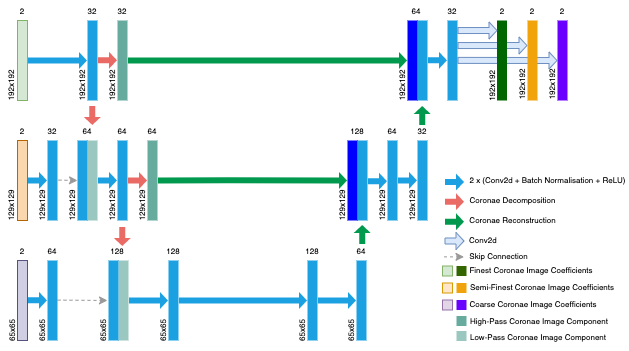}
\caption{Coronae-Net - Architecture here shown with 3 scales for learning the invisible Coronae coefficients in 2D. The input is scale-wise and has 2 channels per scale: the first channel is the visible Coronae coefficient at the corresponding scale and the second channel will be used to predict the unknown invisible Coronae coefficient at the corresponding scale. The Coronae decomposition and reconstruction are used as downsampling and upsampling in the network, respectively.}
\label{image:CoronaeNet}
\end{figure}

Armed with the visible reconstruction $f^\text{vis}$, we now learn the invisible coefficients by a tailored U-Net. As the invisible coefficients are 
in the null space
of the discrete forward operator, there is no need/benefit to include the forward and adjoint operators into the network. 
Our network is based on a U-Net architecture but it works on \emph{Cornae coefficients} and uses the matching \emph{Coronae Decomposition} in lieu of downsampling and the \emph{Coronae Reconstruction} in lieu of upsampling. 
Hence, we term it \emph{Coronae-Net} (\text{CorNet}). The details of  CorNet architecture are depicted in \cref{image:CoronaeNet}.
The inputs of CorNet are visible Coronae coefficients which are generated from the reconstructed visible fully wedge restricted Curvelet coefficients. The visible Coronae coefficients are mapped to a latent representation via a series of convolutions, non-linearities and high/low-pass filters on the down branch of the CorNet. The correlations between the visible and invisible parts of the image are encoded in this latent representation during training and allow prediction of the invisible part of the image. The decoder branch of the CorNet, decodes and upsamples both visible and invisible Coronae coefficients to the highest scale output.

\subsubsection{Rationale for Using Coronae Coefficients as Network Input}
\label{sec:sec:sec:CoronaeInput}
There are two disadvantages to usage of Curvelet coefficients as a network input. Firstly, to preserve the locality of the CNN convolutions (with kernels with localized support, frequently 2 or 3 pixels in each dimension),
they need to be applied in a domain where local convolutions are meaningful. 
A vanilla realisation of a Curvelet based input could entail computing an image domain representation of each individual Curvelet $\Breve{\Psi}^\dagger(C_{j,l}({\bf a}_{j,l}))$ 
which, even when using scale appropriate image sizes, would result in a prohibitive number of image inputs. A better alternative is to use 
all coefficients corresponding to one trapezoidal wedge window (the blue tiles of the Curvelet coefficient plot in \cref{image:BallsCoeffs} (a-c), visualised for one window per scale in \cref{image:CoronaeCurvelet} (a-c)), or their image domain representations in \cref{image:CoronaeCurvelet} (d-f).
However, these are still numerous due to, parabolic scaling induced, angles doubling every second scale. 
Furthermore, the size of the trapezoidal window and hence the corresponding coefficients varies even within the scale which would require additional processing e.g.~zero-padding when working with the coefficients (though not needed when working with their image domain representations \cref{image:CoronaeCurvelet} (d-f)). Finally, it is not clear that training CNNs with small-support kernels would yield good results when working with rather non-smooth and not highly localised (at least not respective to their size) coefficients per window inputs as those in \cref{image:CoronaeCurvelet} (a-c).

We are however, interested in the split into visible and invisible parts rather than in individual (or per window) Curvelet coefficients. This is most economically represented by the Coronae decomposition applied to the visible/invisible parts of the image (see \cref{image:CoronaeCurvelet} (g-i)) which results in a two channel scale-wise input (visible and invisible channel per scale). Moreover, it is apparent from \cref{image:CoronaeCurvelet} that Coronae decomposition is also parsimonious at a higher level feature representation, as it is not reliant on a superposition for final representation unlike Curvelet decomposition. Thus we can reasonably expect it will be easier to learn.

\subsubsection{Coronae-Net}
\label{sec:sec:sec:CoroaneNet}
We propose the following CorNet which we construct with a depth matching the number of scales of the Curvelet decomposition. The input at each scale $j$ consists of 2 channels: the first channel is the Coronae coefficient $Q_j(p^{\textrm{vis}})$ of the visible and the second channel will be used to predict the unknown Coronae coefficient $Q_j(p^{\textrm{inv}})$ of the invisible\footnote{The invisible input is not strictly necessary, however it allows to explicitly inform the network about the structure of the invisible Coronae coefficients, which we deem beneficial due to minimal representation of the visible/invisible and perfect matching of the network scales by the Coronae decomposition.}. At the input stage the latter could be initialised with $\mathbf{0}$ but in practice adding random white noise in training is preferable. 
In the encoder branch, scale by scale the convolutions and nonlinearities extract and correlate the features of the visible and invisible Coronae coefficients to effect the prediction of the invisible Coronae coefficients. The Coronae decomposition is used to pass the low-pass components downward and high-pass components directly to the decoder via the skip connection.
On the up-branch the visible and predicted invisible features are decoded via a symmetric cascade of convolutions and nonlinearities. The Coronae reconstruction is used to assemble the high-pass components passed through the skip connection with the low-pass components obtained from the deeper layer.
We highlight the asymmetry in the CorNet input and output, while we input  the Coronae coefficients scale-wise, the output Coronae coefficients are upsampled (via 0-padding) and returned at the finest scale. Furthermore, the output is multi-headed, as popular in multi-task learning framework, which allows us to assign different weights to coefficients at different scales (and possibly even to visible and invisible channels).

Although the output contains both visible and invisible, the loss function is only applied to the invisible Coronae coefficients and can be written as
\begin{equation}\label{eq:PL}\tag{\textbf{PL}}
    \mathcal{L}(\mathcal{NN}_\text{Cor}(Q^{\text{P}_\text{vis}}), \tilde{Q}^{\text{P}_\text{inv}}) = \sum_{j=j_0}^{j_0+J} \left( \alpha_j\cdot\text{MSE}\left(\mathcal{NN}_\text{Cor}(Q_j^{\text{P}_\text{vis}}),\tilde{Q}_j^{\text{P}_\text{inv}}\right)\right), 
\end{equation}
where the perfect visible Coronae coefficients (original size) $Q_j^{\text{P}_\text{vis}}$ are the input to the CorNet, $\mathcal{NN}_\text{Cor}$, $\tilde{Q}_j^{\text{P}_\text{inv}}$ are the upsampled to the  finest-scale perfect/reference invisible Coronae coefficients, and $\alpha_j$ are some fixed scale dependent weights. The total loss \cref{eq:PL} is summed over the training set of in/output pairs of 
$Q^{\text{P}_\text{vis}}$
and $\tilde{Q}^{\text{P}_\text{inv}}$.

Recall the invisible Curvelet coefficients are in the null space of the  discrete limited-angle PAT Fourier forward operator. The same holds for invisible Coronae coefficients,  c.f.~\Cref{sec:CoronaeVsCurvelets}, 
thus learning the invisible with CorNet from perfect visible coefficients also follows the null space learning paradigm as the scheme presented in \cite{schwab2019deep}, but our chosen representation is essentially optimal for both 
the representation of the visible/invisible parts of the image and preservation of the microlocal properties of the PAT forward operator.
\dontshow{
\begin{equation*}
    Q^{\text{P}_\text{vis}} + \text{P}_{\textbf{ker}(\hat{\textbf{A}}_\angle)}\mathcal{NN}_\text{Cor}(Q^{\text{P}_\text{vis}}),
\end{equation*}
where $\text{P}_{\textbf{ker}(\hat{\textbf{A}}_\angle)}$, the projection on $\textbf{ker}(\hat{\textbf{A}}_\angle)$, extracts the invisible output channel.}

\subsubsection{ResCoronae-Net}
\label{sec:sec:ResCoronaeNet}
The main limitation of the perfect/null space learning is the assumption of availability of perfect visible coefficients. In realistic scenarios the reconstructed visible Coronae coefficients $Q^{\text{vis}}$ will suffer from artefacts introduced by the reconstruction procedure. 
In our case, the Fourier operators contain an error due to interpolation between image and data domain grids.
Therefore, we propose a residual version of the Coronae-Net, 
\emph{ResCoronae-Net} (ResCorNet), to learn an update to both visible and invisible Coronae coefficients. We use the same ${\bf 0}$ or a random noise initialisation for the invisible coefficients as for CorNet, thus the residual connection has no effect for invisible coefficients. We note that we need to upsample our inputs to the finest scale to match the output size before the residual connection (the diagram of ResCoronae-Net can be found in supplementary material Figure SM1). The loss function used for the residual learning penalises errors in both visible and invisible Coronae coefficients
\begin{equation}\tag{\textbf{IPL}}
    \mathcal{L}(\mathcal{NN}_\text{ResCor}(Q^{\text{I}_\text{vis}}), \tilde{Q}^{\text{P}_\text{all}}) = \sum_{j=j_0}^{j_0+J}  \left( \alpha_j\cdot\text{MSE}\left(\mathcal{NN}_\text{ResCor}(Q_j^{\text{I}_\text{vis}}),\tilde{Q}_j^{\text{P}_\text{all}}\right)\right), 
    \label{eq:IPL}
\end{equation}
where the reconstructed imperfect visible Coronae coefficients (original size) $Q_j^{\text{I}_\text{vis}}$ are the input to the ResCorNet, $\mathcal{NN}_\text{ResCor}$, $\tilde{Q}_j^{\text{P}_\text{all}}$ are the upsampled to the finest-scale perfect/reference visible and invisible Coronae coefficients, and $\alpha_j$  some fixed scale dependent weights. Again, the total loss \cref{eq:IPL} is summed over the training set of in/output pairs of $Q^{\text{I}_\text{vis}}$ 
and $\tilde{Q}^{\text{P}_\text{all}}$. 
\dontshow{
\begin{figure}[htbp!]
\centering    
\includegraphics[width=\linewidth]{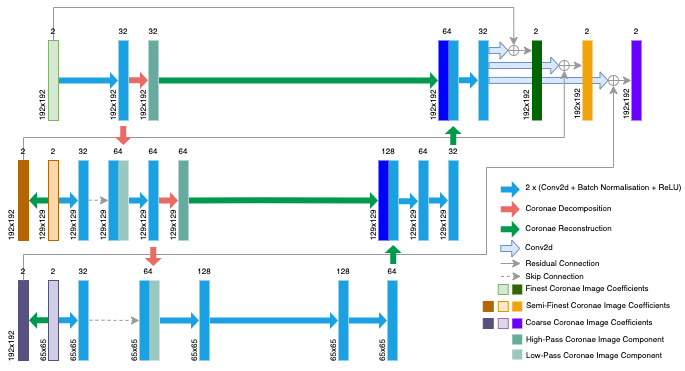}
\caption{ResCoronae-Net - Architecture here shown with 3 scales for correcting the visible and learning the invisible Coronae coefficients in 2D. The input of the ResCoroneNet is scale-wise and upsampled to the finest-scale before residual connections.}
\label{image:ResCoronaeNet}
\end{figure}}

\section{Synthetic Data} 
\label{sec:simulatedexperiments}
In this section, we evaluate the performance of our proposed reconstruction scheme and compare it with various other reconstruction methods. We focus on two learning scenarios: 1) \emph{Perfect Learning}, learns the invisible Coronae coefficients from the given perfect visible Coronae coefficients via CorNet with loss function \cref{eq:PL}; 2) \emph{Imperfect Learning}, learns the update of the visible and the invisible Coronae coefficients, from the imperfect visible Coronae coefficients, constructed from the solution of $\cref{VR}$, via ResCorNet with loss function \cref{eq:IPL}.

\subsection{Experimental Setup}
\label{sec:sec:ExperiementalSetup}
We evaluate our proposed method on an \emph{ellipse data set}. 
The data consists of 3000 synthetic images of superposition of 15 to 20 ellipses with randomly drawn location, orientation and contrast. The images are of resolution $n_{\bx_\perp} \times n_S = 192 \times 192$ assuming a uniform voxel size $h_\bx = 10 \mu\text{m}$. The locations of the ellipses are randomly chosen within the upper half of the image to reduce (while not eliminate completely) the limited sensor effect. The individual images are normalised between 0 and 1 (see \cref{image:PerfectLearnEllipse} (a) for a representative image sample from the ellipse data set). 

We assume homogeneous speed of sound $c=1500$m/s. The pressure time series is recorded every $h_t=66.67$ns by a line sensor with sensitivity angle $\theta_\text{max} = \pi/4$ on top of the domain with image matching resolution i.e.~$h_\bx = 10 \mu\text{m}$,
which we can interpret as number of time-steps $\times$ number of sensors shaped PAT data volume, $272 \times 192$.

We use the fully wedge restricted Curvelet transform with 3 scales and 16 angles in the wedge $W_{\max}$ (at the 2nd coarse scale) for the sparse representation of initial pressure $p_0$. 
The parameters of the transform determine the structure of the networks such as depth (equal to the number of scales, here 3), and the input/output sizes. 

We trained all the networks presented in this section in the same manner to make a fair comparison. 2400 images are used for training, 300 images for validation and 300 for testing. We chose Adam as the optimizer with Xavier initialisation \cite{glorot2010understanding} and batch size of 2. We trained all networks for 200 epochs with 20 epochs early stopping to avoid over-fitting. The initial learning rate is set to $1\cdot10^{-3}$ with a cosine decay. Each network is trained in PyTorch on a single Tesla K40c GPU.

\subsection{Perfect Learning}
\label{sec:sec:PerfectLearning}
We start with an idealised experiment where we learn the invisible Coronae coefficients from the perfect visible Coronae coefficients. The perfect visible/invisible image components and their Coronae coefficients are computed directly via projection e.g.~$p_0^{\text{P}_\text{vis}} = \Breve{\Psi}^\dagger \Breve{\Psi} p_0$, $p_0^{\text{P}_\text{inv}} = p_0 - \Breve{\Psi}^\dagger \Breve{\Psi} p_0$ followed by the Coronae decomposition of $p_0^{\text{P}_\text{vis}},  p_0^{\text{P}_\text{inv}}$ for each image $p_0$ in the training set.

The variation in magnitudes and sizes of Coronae coefficients across scales, suggests use of larger weights for higher scales. 
The choice of $\alpha_\text{F}=2$, $\alpha_\text{SF}=2$ and $\alpha_\text{C}=1$
in \cref{eq:PL} led to the best network performance. We quantitatively compare the results of CorNet with standard variational approaches and U-Net based learned post-processing on the ellipse test data, in terms of MSE, PSNR and SSIM averaged over the test set. 

We compare our results to two reference learned post-processing methods, based on U-Net architecture, trained to remove limited-angle artefacts. 
$\mathcal{NN}_\text{U-Net}(p_0^{\text{P}_\text{vis}})$ takes as an input the perfect visible component $p_0^{\text{P}_\text{vis}}$ and is trained on $(p_0^{\text{P}_\text{vis}},p_0)$ pairs with MSE loss.
$\mathcal{NN}_\text{U-Net}(\tilde{Q}_j^{\text{P}_\text{vis}})$ takes as an input the upsampled to the finest scale (via zero-padding in the Fourier domain) perfect visible Coronae coefficients $\tilde{Q}_j^{\text{P}_\text{vis}}$ (a.k.a.~multichannel input at highest scale only, one channel per scale) and is trained on $(\tilde{Q}_j^{\text{P}_\text{vis}}, \tilde{Q}_j^{\text{P}_\text{all}})$ pairs with MSE loss. 
The diagrams of the reference U-Nets can be found in supplementary material.
\begin{table}[htbp!]
\small
\centering
\caption{Ellipses - Map from perfect visible to invisible. Imaging metrics averaged over the test set.}
\label{tab:quant:Perfectlearn}
\scalebox{0.75}{
\begin{tabular}{c c|c|c|c|c}
    \toprule
    \midrule
    \cmidrule(r){1-1}\cmidrule(rl){2-4}
    \bfseries &  \bfseries MSE & \bfseries PSNR & \bfseries SSIM \\ 
    \cmidrule(r){1-1}\cmidrule(rl){2-4}
    \multicolumn{1}{l}{$p_0^{\text{P}_\text{vis}}$}& $5.8311\cdot10^{-3}\pm2.0110\cdot10^{-3}$ & $22.6029\pm1.5361$ & $0.3612\pm0.0547$ \\
    \multicolumn{1}{l}{${p_0}^{\text{P}_\text{inv}}$}& $1.5827\cdot10^{-2}\pm5.0783\cdot10^{-3}$ & $18.2206\pm1.3726$ & $0.2729\pm0.0574$ \\
    \multicolumn{1}{l}{$\mathcal{NN}_\text{U-Net}(p_0^{\text{P}_\text{vis}})$}& $1.2207\cdot10^{-4}\pm5.4004\cdot10^{-5}$ & $39.5088\pm1.7885$ & $0.9889\pm0.0044$ \\
    \multicolumn{1}{l}{$\mathcal{NN}_\text{U-Net}(\tilde{Q}_j^{\text{P}_\text{vis}})$}& $5.1026\cdot10^{-5}\pm2.4257\cdot10^{-5}$ & $43.3695\pm1.9730$ & $0.9907\pm0.0038$ \\
    \multicolumn{1}{l}{$\mathcal{NN}_\text{Cor}(Q_j^{\text{P}_\text{vis}})$}& $\mathbf{3.4599\cdot 10^{-5}\pm1.4536\cdot 10^{-5}}$ & $\mathbf{44.9678\pm1.7715}$ & $\mathbf{0.9919\pm0.0030}$
\end{tabular}}
\end{table}

\begin{figure}[htbp!]
\centering
  \subfloat[$p_0$]{
  \includegraphics[width=0.23\linewidth,height=0.20\linewidth]{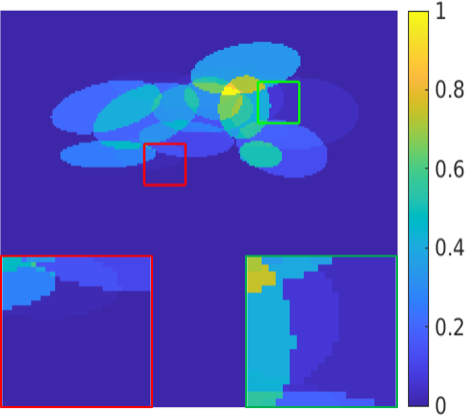}}
  \subfloat[$p_0^{\text{P}_\text{vis}}$]{
  \includegraphics[width=0.23\linewidth,height=0.20\linewidth]{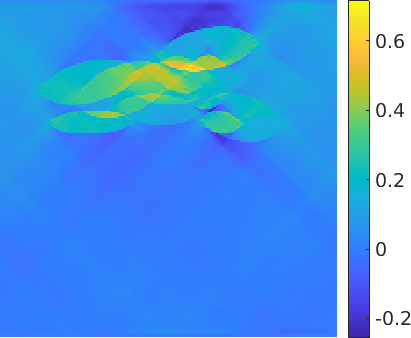}}
  \subfloat[$p_0^{\text{P}_\text{inv}}$]{
  \includegraphics[width=0.23\linewidth,height=0.20\linewidth]{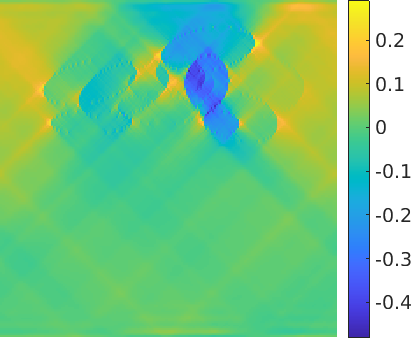}}
  \\
  \subfloat[$\mathcal{NN}_\text{U-Net}(p_0^{\text{P}_\text{vis}})$]{
  \includegraphics[width=0.23\linewidth,height=0.20\linewidth]{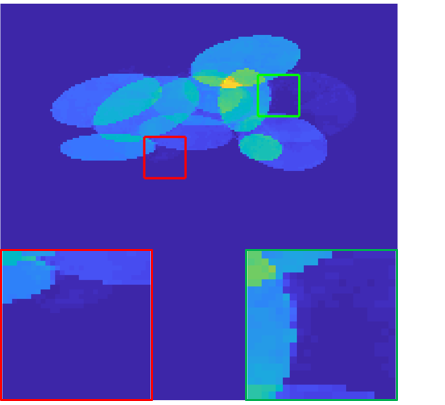}}
   \subfloat[$\mathcal{NN}_\text{U-Net}(\tilde{Q}_j^{\text{P}_\text{vis}})$]{
  \includegraphics[width=0.23\linewidth,height=0.20\linewidth]{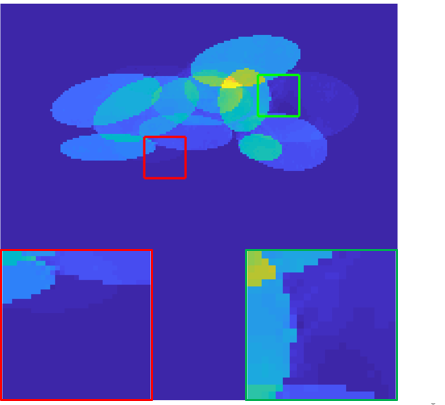}}
  \subfloat[$\mathcal{NN}_\text{Cor}(Q_j^{\text{P}_\text{vis}})$]{
  \includegraphics[width=0.23\linewidth,height=0.20\linewidth]{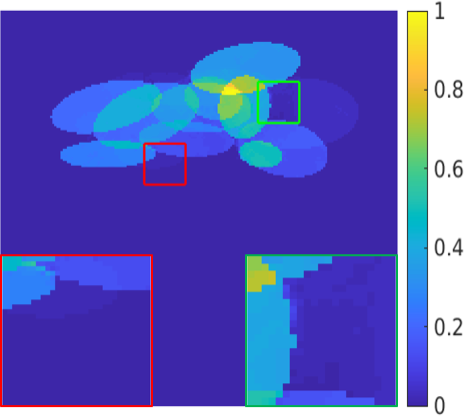}}
  \caption{Ellipses - Visualisation of performance of perfect learning on one test image ($\theta_\text{max} = \pi/4$): (a) initial pressure $p_0$, (b) perfect visible component of $p_0$, (c) perfect invisible component of $p_0$, (d-f) different learned reconstructions.
  }
  \label{image:PerfectLearnEllipse}
\end{figure}
\cref{image:PerfectLearnEllipse} visualises the reconstruction quality for one of the test images. The first row of \cref{image:PerfectLearnEllipse} displays the initial pressure $p_0$ along with its decomposition into perfect visible $p_0^{\text{P}_\text{vis}}$ and perfect invisible $p_0^{\text{P}_\text{inv}}$ parts. The second row of \cref{image:PerfectLearnEllipse} illustrates the learned reconstructions (for plotting only, we restrict the pixel values to [0,1], with values out of this range mapped to the nearest interval end). Although 
all three methods perform very well,
the zoomed-in insets in \cref{image:PerfectLearnEllipse} reveal our proposed scheme yields superior reconstruction in terms of invisible boundary (green inset), ellipse shapes and faint contrasts (red inset). 
On average, $\mathcal{NN}_\text{Cor}(Q_j^{\text{P}_\text{vis}})$ achieves lowest MSE, highest PSNR and highest SSIM in \cref{tab:quant:Perfectlearn} which is consistent with better, in eye ball metric, reconstructions produced by the CorNet\footnote{The invisible Coronae coefficients $\tilde{Q}_j^{\text{L}_\text{inv}}$ learned using CorNet can be found in supplementary material.}. However, we observe all methods in (d-f) fail to recover the faint ellipse on top of even fainter large ellipse and overlapping with ellipses with higher contrasts (highlighted in green inset). The low contrast boundary of the faint ellipse is difficult to pick up for the networks. We encountered a similar scenario in \cref{image:CoronaeDRImage}, where the boundary of the faint blue disc is almost invisible in the high-pass component $\mathbb{H} (p_0)$ while the boundaries of the ellipses with higher contrast are still perceptible. 

\subsection{Imperfect Learning}
\label{sec:sec:ImperfectLearning}
Next, we consider the more practical scenario of correcting the visible and learning the invisible Coronae coefficients on the ellipse data set from limited-angle PAT data with additive white noise with standard deviation $\sigma = 2.5\cdot10^{-4}$. We reconstruct the visible Curvelet coefficients $\tilde{f}^\text{vis}$ solving \cref{VR} for regularisation parameter $\tau = 2.5\cdot10^{-4}$ performing 50 FISTA iterations. From the point of view of the noise level this corresponds to an early stopping, however the convergence of FISTA has already slowed down from the theoretical rate $\mathcal O(1/k^2)$ due to interpolation errors in the forward operator, thus 50 FISTA iterations constitute a reasonable computing time / accuracy trade-off and provide a good starting point for learning the correction. \dontshow{$K_\text{max}=50$} This scenario employs ResCorNet with the scale dependent loss weights $\alpha_\text{F}=2$, $\alpha_\text{SF}=2$ and $\alpha_\text{C}=1$ in \cref{eq:IPL}. Again, we compare the results obtained with ResCorNet against and residual variants of U-Net learned post-processing methods on the test subset of the ellipse data. 

We compare our results to five reference reconstruction methods applied to noisy limited-angle data $g_\angle$. $p_0^\text{Linear} = \hat{\textbf{A}}_{\angle}^{-1} g_\angle$: linear inversion. 
${p_0}^{\text{I}_\text{vis}}_{\ell_1}$: recovery of the sparse representation of the visible part of $p_0$ in fully wedge restricted Curvelet frame via \cref{VR}.
${p_0}^{\text{I}_\text{vis}}_{\text{TV}+}$: total variation regularised solution with non-negativity constraint \cite{arridge2016accelerated}. 
And two learned post-processing methods based on ResU-Net architecture trained to remove artefacts due to limited sensitivity angle and errors in the visible part of the image/Coronae coefficients.
$\mathcal{NN}_\text{ResU-Net}(p_0^{\text{I}_\text{vis}})$: 
takes as an input the imperfect visible coefficients $p_0^{\text{I}_\text{vis}}$ and is trained on $(p_0^{\text{I}_\text{vis}}, p_0)$ pairs with MSE loss. 
$\mathcal{NN}_\text{ResU-Net}(\tilde{Q}_j^{\text{I}_\text{vis}})$: 
takes as an input the (upsampled to the finest scale via zero-padding in the Fourier domain) imperfect visible Coronae coefficients $\tilde{Q}_j^{\text{I}_\text{vis}}$ and is trained on $(\tilde{Q}_j^{\text{I}_\text{vis}}, \tilde{Q}_j^{\text{P}_\text{all}})$ pairs with MSE loss. The diagrams of the reference ResU-Nets can be found in supplementary material.
\begin{table}[htbp!]
\small
\centering
\caption{Ellipses - Map from imperfect visible to corrected visible and learned invisible. Imaging metrics averaged over test set. Note, in the top two rows, perfect visible component $p_0^{\text{P}_\text{vis}}$ is used as a ground truth while, in the remaining rows, we use the original image $p_0$.}
\label{tab:quant:Imperfectlearn}
\scalebox{0.75}{
\begin{tabular}{c c|c|c|c|c}
    \toprule
    \midrule
    \cmidrule(r){1-1}\cmidrule(rl){2-4}
    \bfseries &  \bfseries MSE & \bfseries PSNR & \bfseries SSIM \\ 
    \cmidrule(r){1-1}\cmidrule(rl){2-4}
    \multicolumn{1}{l}{${p_0}^{\text{Linear}}$}& $1.5031\cdot10^{-4}\pm6.0914\cdot10^{-5}$ & $38.5706\pm1.7347$ & $0.8296\pm0.0460$  \\
    \multicolumn{1}{l}{${p_0}^{\text{I}_\text{vis}}_{\ell_1}$}& $9.5904\cdot{10}^{-5}\pm3.6300\cdot10^{-5}$ & $40.4581\pm1.5361$ & $0.9577\pm0.0090$  \\
    \cmidrule(r){1-1}\cmidrule(rl){2-4}
    \multicolumn{1}{l}{${p_0}^{\text{Linear}}$}& $6.0860\cdot10^{-3}\pm2.2048\cdot10^{-3}$ & $22.4308\pm1.5593$ & $0.3284\pm0.0436$  \\
    \multicolumn{1}{l}{${p_0}^{\text{I}_\text{vis}}_{\ell_1}$}& $5.9660\cdot{10}^{-3}\pm2.1752\cdot10^{-3}$ & $22.5209\pm1.5700$ & $0.3463\pm0.0580$  \\
    \multicolumn{1}{l}{${p_0}^{\text{I}_\text{vis}}_\text{TV+}$}& $8.6550\cdot10^{-5}\pm4.4774\cdot10^{-5}$ & $41.0863\pm1.9422$ & $\mathbf{0.9861\pm0.0059}$  \\
    \multicolumn{1}{l}{$\mathcal{NN}_\text{ResU-Net}(p_0^{\text{I}_\text{vis}})$}& $1.4197\cdot10^{-4}\pm6.3716\cdot10^{-5}$ & $38.8494\pm1.7685$ & $0.9586\pm0.0133$ \\
    \multicolumn{1}{l}{$\mathcal{NN}_\text{ResU-Net}(\tilde{Q}_j^{\text{I}_\text{vis}})$}& $9.4125\cdot10^{-5}\pm3.7918\cdot10^{-5}$ & $40.5950\pm1.7027$ & $0.9705\pm0.0074$ \\
    \multicolumn{1}{l}{$\mathcal{NN}_\text{ResCor}(Q_j^{\text{I}_\text{vis}})$}& $\mathbf{8.1092\cdot10^{-5}\pm3.0781\cdot10^{-5}}$ & $\mathbf{41.2081\pm1.6141}$ & $0.9755\pm0.0069$ 
\end{tabular}}
\end{table}

\begin{figure}[htbp!]
\centering
  \subfloat[$p_0^\text{Linear}$]{
  \includegraphics[width=0.23\linewidth,height=0.20\linewidth]{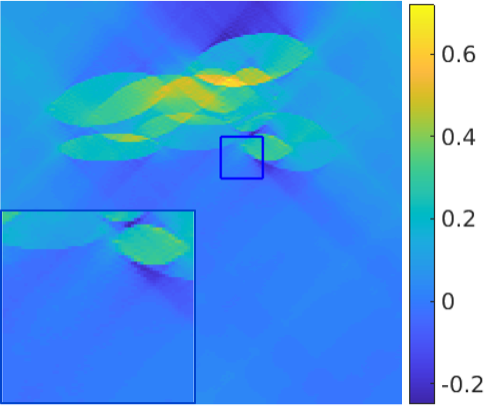}}
  \subfloat[${p_0}^{\text{I}_\text{vis}}_{\ell_1}$]{
  \includegraphics[width=0.23\linewidth,height=0.20\linewidth]{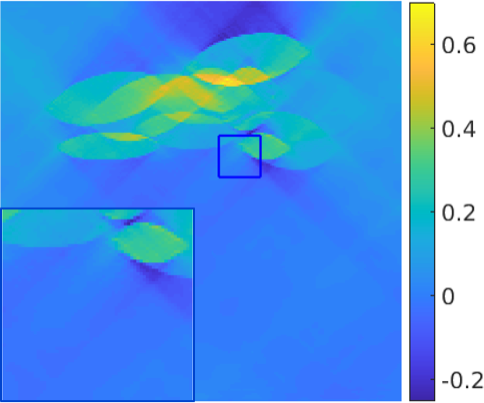}}
  \subfloat[${p_0}^{\text{I}_\text{vis}}_\text{TV+}$]{
  \includegraphics[width=0.23\linewidth,height=0.20\linewidth]{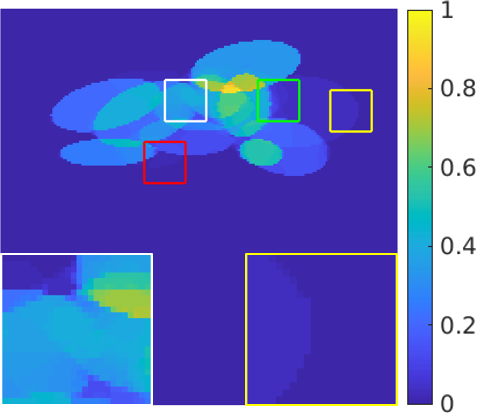}}
  \\
  \subfloat[$\mathcal{NN}_\text{ResU-Net}(p_0^{\text{I}_\text{vis}})$]{
  \includegraphics[width=0.23\linewidth,height=0.20\linewidth]{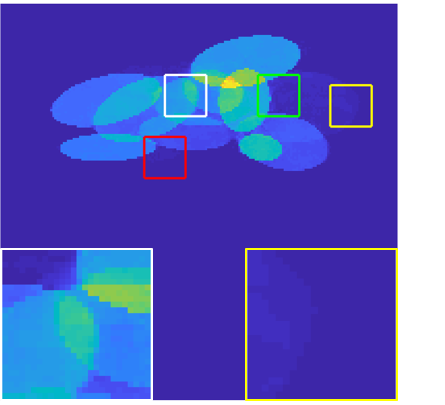}}
   \subfloat[$\mathcal{NN}_\text{ResU-Net}(\tilde{Q}_j^{\text{I}_\text{vis}})$]{
  \includegraphics[width=0.23\linewidth,height=0.20\linewidth]{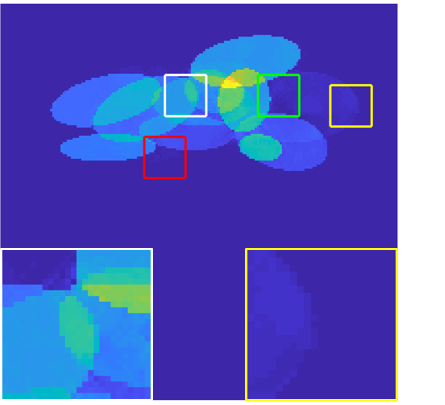}}
  \subfloat[$\mathcal{NN}_\text{ResCor}(Q_j^{\text{I}_\text{vis}})$]{
  \includegraphics[width=0.23\linewidth,height=0.20\linewidth]{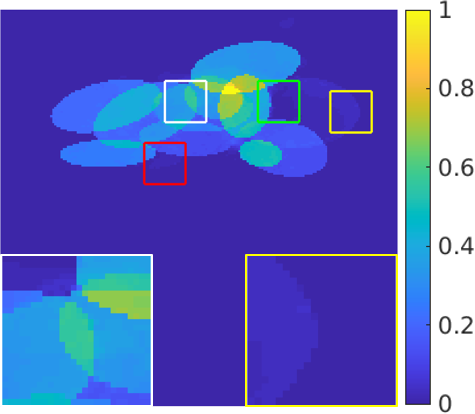}}
  \caption{Ellipses - Visualisation of performance of the imperfect learning on noisy limited-angle data $g_\angle$, $\theta_\text{max} = \pi/4$, for one test image: (a) linear inversion, (b) reconstructed imperfect visible component via \cref{VR}, (c) total variation regularised solution with non-negativity constraint, (d-f) different learned reconstructions. 
  }
  \label{image:ImperfectLearnEllipse}
\end{figure}

\begin{table}[htbp!]
\small
\centering
\caption{Ellipses - Timing and resources for imperfect learning experiment.}
\label{tab:quant:ImperfectlearnTime}
\scalebox{0.75}{
\begin{tabular}{c c|c|c|c|c}
    \toprule
    \midrule
    \cmidrule(r){1-1}\cmidrule(rl){2-4}
    \bfseries &  \bfseries Train & \bfseries Execution & \bfseries Parameter\\ 
    \cmidrule(r){1-1}\cmidrule(rl){2-4}
    \multicolumn{1}{l}{${p_0}^{\text{Linear}}$}& \dontshow{3m32s} - & 1m52s & 0 \\
    \multicolumn{1}{l}{${p_0}^{\text{I}_\text{vis}}_{\ell_1}$}&  \dontshow{1h02m} - & 16m23s& 1  \\
    \multicolumn{1}{l}{${p_0}^{\text{I}_\text{vis}}_\text{TV+}$}&  \dontshow{54m43s} - & 14m12s & 1 \\
    \multicolumn{1}{l}{$\mathcal{NN}_\text{ResU-Net}(p_0^{\text{I}_\text{vis}})$}& 6h05m & $\sim16$s & 467554\\
    \multicolumn{1}{l}{$\mathcal{NN}_\text{ResU-Net}(\tilde{Q}_j^{\text{I}_\text{vis}})$}& 8h12m & $\sim24$s & 468838 \\
    \multicolumn{1}{l}{$\mathcal{NN}_\text{ResCor}(Q_j^{\text{I}_\text{vis}})$}& 18h48m & $\sim24$s & 1001670
\end{tabular}}
\end{table}

The image quality measures averaged over the 300 test images are listed in \cref{tab:quant:Imperfectlearn}, specifically MSE, PSNR and SSIM with respect to the provided ground-truth images. The reconstruction quality is illustrated on an example from the test set in \cref{image:ImperfectLearnEllipse}. Despite ${p_0}^{\text{I}_\text{vis}}_{\text{TV}+}$ achieving highest SSIM, the reconstructed boundaries of the ellipses highlighted in the white inset in \cref{image:ImperfectLearnEllipse} (c) are incomplete (total variation apparently fails to completely recover the invisible boundaries) and considerably worse than for the learning based reconstructions (d-f). We note that the linear reconstruction ${p_0}^{\text{Linear}}$ in \cref{image:ImperfectLearnEllipse} (a) and ${p_0}^{\text{I}_\text{vis}}_{\ell_1}$ in \cref{image:ImperfectLearnEllipse} (b) look similar with a small difference in contrast. The linear inversion results in slightly less clear ellipse boundaries and noisier image than the non-linear $\ell_1$ reconstruction; see blue inset in \cref{image:ImperfectLearnEllipse} (a,b). 
We recall, that the linear inversion here is able to recover the visible component due to limited sensitivity angle (and even limited sensor) completely, thus any difference should be due to higher robustness of the non-linear reconstruction to noise.
This is consistent with the latter's lower MSE, higher PSNR and higher SSIM compared to the linear inversion. The second row in \cref{image:ImperfectLearnEllipse} (restricted to range [0,1]) shows the results of the post-processing with ResU-Nets and ResCorNet, respectively. Although the ResU-Net based methods in \cref{image:ImperfectLearnEllipse} (d,e) estimate the invisible boundaries reasonably well they exhibit some high frequency noise. Upon closer inspection (white and yellow insets in \cref{image:ImperfectLearnEllipse}) and quantitative analysis in \cref{tab:quant:Imperfectlearn}, our proposed ResCorNet seems to strike the best balance between the recovery of invisible boundaries while preserving the piece-wise constant interior of the ellipses. Similarly, as in the perfect learning scenario, all three methods in (d-f) fail to recover the faint green framed ellipse.

The timing and resources are listed in \cref{tab:quant:ImperfectlearnTime} (Note, ${p_0}^{\text{I}_\text{vis}}_{\ell_1}$ and ${p_0}^{\text{I}_\text{vis}}_{\text{TV+}}$ are computed using Matlab Parallel Computing Toolbox\footnote{https://www.mathworks.com/products/parallel-computing.html}). Clearly $\mathcal{NN}_\text{ResU-Net}(p_0^{\text{I}_\text{vis}})$ is fastest to train and execute, while $\mathcal{NN}_\text{ResCor}(Q_j^{\text{I}_\text{vis}})$ requires longest training time. The main reason is due to ResCorNet's larger number of trainable parameters: 1) the preceding convolutions at each scale are paramount to extract the features from the scale-wise Coronae coefficients input; 2) the Coronae coefficients sizes are determined by the partition of unity filters underpinning Curvelet decomposition, which result in larger sizes of layer inputs at each but the finest scale than those obtained from max-pooling. It is noteworthy, that at prediction stage, the approaches learning the Coronae coefficients $\mathcal{NN}_\text{ResU-Net}(\tilde{Q}_j^{\text{I}_\text{vis}})$ and $\mathcal{NN}_\text{ResCor}(Q_j^{\text{I}_\text{vis}})$ only 
have a slight overhead 
compared to $\mathcal{NN}_\text{ResU-Net}(p_0^{\text{I}_\text{vis}})$, and are much faster than the variational reconstruction approaches.

\subsection{Generalisation}
\label{sec:Generalisation}
\begin{figure}[htbp!]
\centering
  \subfloat[$\mathcal{NN}_\text{ResU-Net}(p_0^{\text{I}_\text{vis}})$]{
  \includegraphics[width=0.23\linewidth,height=0.20\linewidth]{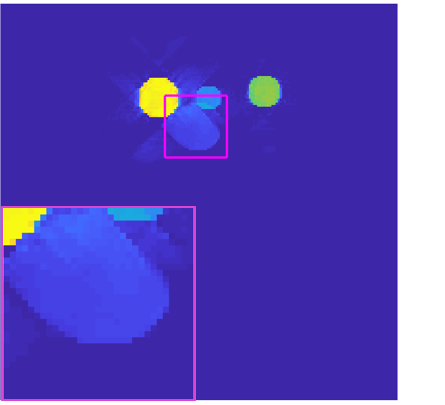}}
   \subfloat[$\mathcal{NN}_\text{ResU-Net}(\tilde{Q}_j^{\text{I}_\text{vis}})$]{
  \includegraphics[width=0.23\linewidth,height=0.20\linewidth]{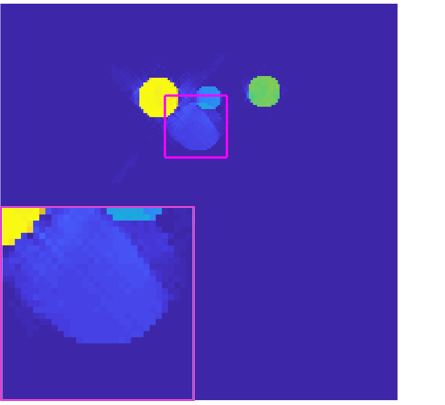}}
  \subfloat[$\mathcal{NN}_\text{ResCor}(Q_j^{\text{I}_\text{vis}})$]{
  \includegraphics[width=0.23\linewidth,height=0.20\linewidth]{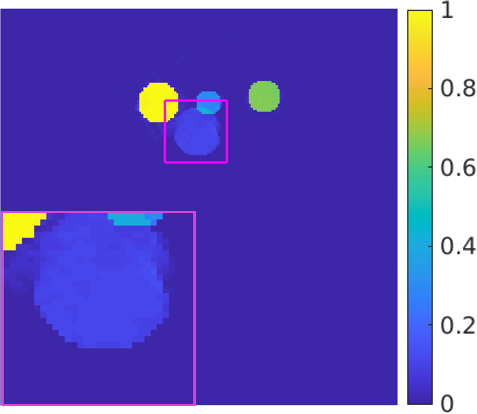}}
  \caption{4-disk Generalisation - Visualisation of generalisation performance of correcting the visible and learning the invisible from ellipse data to 4-disk phantom ($\theta_\text{max} = \pi/4$): (a-c) different learned reconstructions.}
  \label{image:ImperfectLearnBall}
\end{figure}

We now consider the generalisation properties of (Res)CorNet. To this end we chose a 4-disk phantom depicted in \cref{image:ImperfectLearnBall}. The 4-disk phantom is not in the ellipse data set (in the sense how the set is generated) while, a disk being a special case of an ellipse, has similar features 
such as convexity and boundaries of $C^2$ type. In this experiment, the visible Curvelet coefficients are obtained solving $\cref{VR}$ for $\tau = 2.5\cdot10^{-4}$ with \dontshow{$K_\text{max} = 50$} 50 FISTA iterations. Again the white noise with standard deviation $\sigma = 2.5\cdot10^{-4}$ is added to the corresponding limited-angle data with $\theta_\text{max} = \pi/4$. We apply the ResCorNet trained on the ellipse data set 
to the 4-disk phantom. This scenario corresponds to training on a complicated data set and generalising our model to a less complex image with some shared and some unseen characteristics. We note that the training set contains many ellipses at least partially outside the north sector of the domain (inside which the limited sensor has no bearing on the reconstruction) while our 4-disk phantom lies strictly inside. The main limitation of the ResU-Net based reconstructions in \cref{image:ImperfectLearnBall} (a,b) are streak like artefacts and failure to recover the invisible boundary of the large blue disk highlighted in the magenta inset. 
Our proposed reconstruction method using ResCorNet in \cref{image:ImperfectLearnBall} (c) provides a notable improvement in the definition of the invisible boundary 
as well as the reduction in streak like background artefacts, which is consistent with its lowest MSE, highest PSRN and SSIM in \cref{tab:quant:generaliseBall}. 
However, even in (c) parts of the invisible boundary of the faint large blue disk are missing (both left and right sides). We believe these artefacts could be due to the training set  containing images of many overlapping ellipses, which means that examples of boundaries against the background are rare, only present for exterior ellipses.
Furthermore, we note that the recovered shape has some resemblance with an ellipse. Both observations testify to limits of generalisation. The remaining disks are almost perfectly reconstructed by all methods. Overall, our proposed method while not without limitations is clearly superior in this generalisation scenario both in eye ball metric and the quantitative analysis in \cref{tab:quant:generaliseBall}.
\begin{table}[htbp!]
\small
\centering
\caption{4-disk - Generalisation - Ellipse learned map from imperfect visible to corrected visible and learned invisible.}
\label{tab:quant:generaliseBall}
\scalebox{0.75}{
\begin{tabular}{c c|c|c|c|c}
    \toprule
    \midrule
    \cmidrule(r){1-1}\cmidrule(rl){2-4}
    \bfseries &  \bfseries MSE & \bfseries PSNR & \bfseries SSIM \\ 
    \cmidrule(r){1-1}\cmidrule(rl){2-4}
    \multicolumn{1}{l}{$\mathcal{NN}_\text{ResU-Net}(p_0^{\text{I}_\text{vis}})$}& $2.0000\cdot10^{-4}$ & 36.9776 & 0.9398 \\
    \multicolumn{1}{l}{$\mathcal{NN}_\text{ResU-Net}(\tilde{Q}_j^{\text{I}_\text{vis}})$}& $1.0000\cdot10^{-4}$ & 39.8984 & 0.9446 \\
    \multicolumn{1}{l}{$\mathcal{NN}_\text{ResCor}(Q_j^{\text{I}_\text{vis}})$}& $\mathbf{2.9599\cdot 10^{-5}}$ & $\mathbf{45.2885}$ & $\mathbf{0.96866}$
\end{tabular}}
\end{table}

\section{Realistic Data} 
\label{sec:RealDataApplication}
\begin{figure}[htbp!]
\centering
  \subfloat[$p_0$]{
  \includegraphics[width=0.23\linewidth,height=0.20\linewidth]{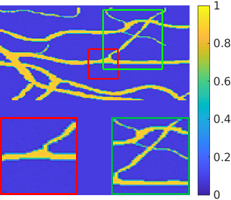}}
  \subfloat[${p_0}^{\text{I}_\text{vis}}_{\ell_1}$]{
  \includegraphics[width=0.23\linewidth,height=0.20\linewidth]{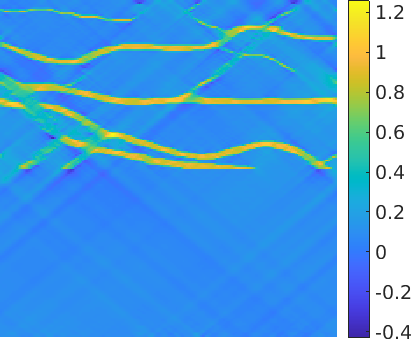}}
  \subfloat[${p_0}^{\text{I}_\text{vis}}_\text{TV+}$]{
  \includegraphics[width=0.23\linewidth,height=0.20\linewidth]{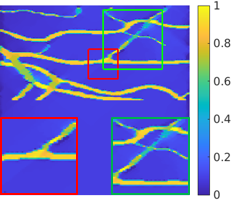}}
  \\
  \subfloat[$\mathcal{NN}_\text{ResU-Net}(p_0^{\text{I}_\text{vis}})$]{
  \includegraphics[width=0.23\linewidth,height=0.20\linewidth]{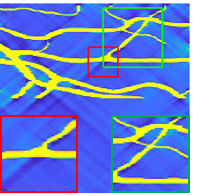}}
   \subfloat[$\mathcal{NN}_\text{ResU-Net}(\tilde{Q}_j^{\text{I}_\text{vis}})$]{
  \includegraphics[width=0.23\linewidth,height=0.20\linewidth]{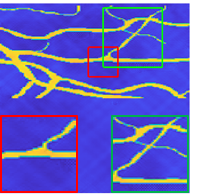}}
  \subfloat[$\mathcal{NN}_\text{ResCor}(Q_j^{\text{I}_\text{vis}})$]{
  \includegraphics[width=0.23\linewidth,height=0.20\linewidth]{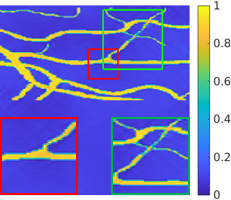}}
  \caption{Vessels - Visualisation of performance of correcting the visible and learning the invisible 
  on noisy limited-angle data $g_\angle$, $\theta_\text{max} =  2\pi/9$, for one test image: (a) initial pressure $p_0$ (ground truth), (b) imperfect visible component reconstruction via \cref{VR}, (c) total variation regularised solution with non-negativity constraint, (d-f) different learned reconstructions.
  }
  \label{image:ImperfectLearnVessel}
\end{figure}

In this section we evaluate performance of our method
on a realistic \dontshow{an \emph{in-vivo}} vessel data set. Our data set of vessel images is generated by randomly cropping images from the DRIVE data set for retinal vessel segmentation\footnote{https://drive.grand-challenge.org/}. The re-sampled to $192 \times 96$ and normalised to [0,1] vessel images constitute the top half of the test image, with bottom half all 0, resulting in test images of size $192 \times 192$, see \cref{image:ImperfectLearnVessel}. We assume homogeneous speed of sound $c=1500$m/s. The pressure time series is recorded every $h_t=66.67$ns by a line sensor, placed on top of the domain, with image matching resolution i.e.~$h_\bx = 10 \mu\text{m}$ resulting in a $272 \times 192$ PAT data volume. 
The 
sensitivity angle $\theta_\text{max} = 2\pi/9$ is slightly smaller (and so is the visibility cone) than in the previous examples. 
2000 images are used for training, 200 for validation and 200 for testing.

We perform the visible reconstruction \cref{VR} in fully wedge restricted Curvelet frame with 3 scales and 12 angles in the visible wedge (at the 2nd coarsest scale)
for the regularisation parameter $\tau = 1\cdot10^{-4}$ with 50 FISTA iterations. 
As we are in the imperfect learning scenario, we train the matching ResCorNet to minimise the loss \cref{eq:IPL} with weights $\alpha_\text{F}=2$, $\alpha_\text{SF}=2$ and $\alpha_\text{C}=1$
in the same regime as outlined in \Cref{sec:simulatedexperiments}.

The Coronae coefficient based learning
$\mathcal{NN}_\text{ResU-Net}(\tilde{Q}_j^{\text{I}_\text{vis}})$ and $\mathcal{NN}_\text{ResCor}(Q_j^{\text{I}_\text{vis}})$ outperforms image based learning $\mathcal{NN}_\text{ResU-Net}(p_0^{\text{I}_\text{vis}})$ as clearly visible in reconstructions in \cref{image:ImperfectLearnVessel}. $\mathcal{NN}_\text{ResU-Net}(p_0^{\text{I}_\text{vis}})$ suffers from strong limited-angle artefacts while ${p_0}^{\text{I}_\text{vis}}_{\text{TV+}}$ fails to recover the invisible vessel edges completely. Zooming in \cref{image:ImperfectLearnVessel} (red and green insets) corroborates that ResCorNet produces smooth background, sharp and detailed vessel edges and well matched contrasts. Some of these are also present in the other reconstructions but ResCorNet is unique in achieving all these goals simultaneously. This is reminiscent of ResCorNet performance for the ellipse data set (see \Cref{sec:sec:ImperfectLearning}). 

The quantitative analysis is summarised in \cref{tab:quant:ImperfectlearnVessel}. Most notably, we observe a large performance leap across all the measures between the methods learning Coronae coefficients and those learning images which reaffirms the benefits of Coronae representation in this context.
\begin{table}[htbp!]
\small
\centering
\caption{Vessels - Map from imperfect visible to corrected visible and learned invisible. Imaging metrics averaged over test set. Note, in the top two rows, perfect visible component $p_0^{\text{P}_\text{vis}}$ is used as a ground truth while, in the remaining rows, we use the original image $p_0$.
}
\label{tab:quant:ImperfectlearnVessel}
\scalebox{0.75}{
\begin{tabular}{c c|c|c|c|c}
    \toprule
    \midrule
    \cmidrule(r){1-1}\cmidrule(rl){2-4}
    \bfseries &  \bfseries MSE & \bfseries PSNR & \bfseries SSIM \\ 
    \cmidrule(r){1-1}\cmidrule(rl){2-4}
     \multicolumn{1}{l}{${p_0}^{\text{Linear}}$}& $1.3135\cdot10^{-3}\pm5.0308\cdot10^{-4}$ & $29.1241\pm1.6788$ & $0.8836\pm0.0529$  \\
    \multicolumn{1}{l}{${p_0}^{\text{I}_\text{vis}}_{\ell_1}$}& $2.5338\cdot{10}^{-4}\pm1.4832\cdot10^{-4}$ & $36.5491\pm2.2176$ & $0.9537\pm0.0330$  \\
    \cmidrule(r){1-1}\cmidrule(rl){2-4}
    \multicolumn{1}{l}{${p_0}^{\text{I}_\text{vis}}_{\ell_1}$}& $1.0201\cdot10^{-2}\pm6.6000\cdot10^{-3}$ & $20.7758\pm2.8440$ & $0.7947\pm0.1086$  \\
    \multicolumn{1}{l}{${p_0}^{\text{I}_\text{vis}}_\text{TV+}$}& $4.3110\cdot10^{-3}\pm3.0828\cdot10^{-3}$ & $24.7764\pm3.3682$ & $0.9398\pm0.0292$  \\
    \multicolumn{1}{l}{$\mathcal{NN}_\text{ResU-Net}(p_0^{\text{I}_\text{vis}})$}& $4.5594\cdot10^{-2}\pm1.7357\cdot10^{-2}$ & $13.7066\pm1.6040$ & $0.6562\pm0.1104$ \\
    \multicolumn{1}{l}{$\mathcal{NN}_\text{ResU-Net}(\tilde{Q}_j^{\text{I}_\text{vis}})$}& $\mathbf{7.4186\cdot10^{-4}\pm6.6170\cdot10^{-4}}$ & $32.3976\pm2.9230$ & $0.9540\pm0.0717$ \\
    \multicolumn{1}{l}{$\mathcal{NN}_\text{ResCor}(Q_j^{\text{I}_\text{vis}})$}& $7.5803\cdot10^{-4}\pm7.0400\cdot10^{-4}$ & $\mathbf{32.4036\pm3.0706}$ & $\mathbf{0.9611\pm0.0714}$ 
\end{tabular}}
\end{table}

\section{Conclusion}
\label{sec:Conclusion}
The paper rigorously builds a framework where images are nearly optimally sparsely represented in a fully wedge restricted Curvelet frame allowing splitting into invisible/visible components, which are optimally matched with the null space/its complement of the discrete forward operator. All computations can be performed efficiently in the Fourier domain. The network architecture working with Coronae coefficients of the visible and invisible image components is carefully designed to match both the visible/invisible split and the multiscale decomposition induced by the frame. 
Even, the sole learning of Coronae coefficients, as evidenced by our reference U-Net working on Coronae coefficients, already results in a substantial improvement on an image based learned post-processing using a standard U-Net.
The resulting scheme implements the
\textit{reconstructing the visible and learning the invisible} strategy which
decouples the forward model from the training process at no detriment to reconstruction accuracy. This results in a stark performance benefit and a better interpretability compared to the model based learning which is considered the gold standard (at least in terms of image quality) among learned reconstructions.


We believe that our framework for this particular class of limited-view problems in photoacoustic tomography will further interest in applying similar ideas to other inverse problems, in particular application to the limited-angle parallel X-ray CT is immediate. As, with all learned methods, performance of our methods strongly depends on appropriateness and quality of the training data. Finally, we note that extension of our method to 3D is conceptually straightforward and, as no expensive forward/adjoint operators are evaluated in training, it is computationally feasible.

\appendix
\section{Adjoint PAT Fourier Operator}
\label{appendixlabel1}
By the definition, the adjoint PAT Fourier operator \cref{eq:FrequencyEquationAdj} satisfies the equality 
\begin{equation}\label{eq:defAdjoint}
\langle \hat{g},\hat{\textbf{A}}\hat p_0 \rangle = \langle \hat{\textbf{A}}^*\hat{g},\hat p_0 \rangle
\end{equation}
for any real and even w.r.t.~first variable functions $\hat g(\omega/c,\bk_S)$, $\hat p(\omega/c, \bk_S)$, $(\omega/c,\bk_S) \in \mathcal R_\textbf{A} \subset \mathbb R^d$ and $\hat p_0(\bk_\perp, \bk_S)$, $(\bk_\perp, \bk_S) \in \mathbb R^{d}$ with sufficient regularity and decay (essentially in Schwarz space over $\mathbb R^d$) so that all the operations are well defined and $\langle \cdot, \cdot \rangle$ is the standard $L_2$ scalar product which we restrict to $\mathcal R_\textbf{A}$ for functions over $(\omega/c,\bk_S) \in \mathcal R_\textbf{A}$.

Substituting the action of the forward operator $\hat{\textbf{A}}$ on $\hat p_0$, $\hat p(\omega/c, \bk_S) = \hat{\textbf{A}} \hat p_0(\bk_\perp, \bk_S)$, on the left hand side of
\cref{eq:defAdjoint}, we obtain
\begin{align}
 \label{eq:lhsAdjoint} \langle \hat{g},\hat{\textbf{A}} \hat p_0 \rangle & = \langle \hat{g}(\omega/c, \bk_S),\hat p(\omega/c, \bk_S) \rangle \\
 \nonumber   & = 2 \int\limits_{\mathbb{R}_{\geq 0}} \int\limits_{\mathbb{R}^{d-1}}  \hat{g} (\omega/c, \bk_S) \frac{\omega/c}{\sqrt{(\omega/c)^2-{|\mathbf{k}_S}|^2}} \hat p_0\left(\sqrt{ (\omega/c)^2- |\mathbf{k}_S}|^2, \bk_S\right) d(\omega/c) d{\bk_S} 
\end{align}
where we used the even symmetry in the first variable of $\hat g(\omega/c,\bk_S)$, $\hat p(\omega/c,\bk_S)$, which in turn implies even symmetry of $\hat p_0(\bk_\perp, \bk_S)$ (used later), to restrict the 
the first integral to non-negative real axis $\mathbb R_{\geq 0}$.

In the next step we make use of change of variables induced by the dispersion relation in the wave equation \cref{eq:map:k}, $(\bk_\perp, \bk_S) \leftarrow \left( \sqrt{(\omega/c)^2 - |\bk_S|^2}, \bk_S \right), \, \bk_\perp \geq 0$,
assumed continued with even symmetry for $\bk_\perp < 0$. The Jacobian of the change of coordinates yields
\begin{equation*}
d \bk_\perp= \frac{\omega/c}{\sqrt{(\omega/c)^2-|\mathbf{k}_S|^2}}d(\omega/c).
\end{equation*}
Executing the change of coordinates in the last line of \cref{eq:lhsAdjoint}, the factor gets absorbed into the volume element and we are left with
\begin{align*}
    \langle \hat{g},\hat{\textbf{A}}p_0 \rangle 
    & = \iint\limits_{\mathbb R^d} \hat{g} (|\bk|, \bk_S)\hat p_0(\bk_{\perp},\bk_S) d\bk_{\perp} d{\bk_S}\\
    & = \langle \hat g( |\bk|, \bk_S ) , \hat p_0(\bk_\perp, \bk_S) \rangle,
\end{align*}
where the factor 2 is accounted for by the extension of the integration domain of the product of even functions from $\mathbb{R}_{\geq 0}$ back to $\mathbb{R}$. 
Comparing the above equation to the right hand side of the defining equality \cref{eq:defAdjoint}, we immediately read out the action of the adjoint operator on $\hat g(\omega/c , \bk_S)$ as
\begin{align}
\hat{\textbf{A}}^* \hat g(\omega/c , \bk_S) &=  \hat g(|\bk| , \bk_S).
\end{align}

\bibliographystyle{siamplain}
\bibliography{references}

\end{document}